\documentclass[12pt]{article}

\usepackage{amsmath}
\usepackage{amssymb}

\allowdisplaybreaks

\begin{document}

\baselineskip=18.8pt plus 0.2pt minus 0.1pt

%%%%%%%%%%% Private Macros %%%%%%%%%%%%%
\makeatletter

\@addtoreset{equation}{section}
\renewcommand{\theequation}{\thesection.\arabic{equation}}
\renewcommand{\thefootnote}{\fnsymbol{footnote}}
\newcommand{\beq}{\begin{equation}}
\newcommand{\eeq}{\end{equation}}
\newcommand{\bea}{\begin{eqnarray}}
\newcommand{\eea}{\end{eqnarray}}
\newcommand{\nn}{\nonumber\\}
\newcommand{\hs}[1]{\hspace{#1}}
\newcommand{\vs}[1]{\vspace{#1}}
\newcommand{\Half}{\frac{1}{2}}
\newcommand{\p}{\partial}
\newcommand{\ol}{\overline}
\newcommand{\wt}[1]{\widetilde{#1}}
\newcommand{\ap}{\alpha'}
\newcommand{\bra}[1]{\left\langle  #1 \right\vert }
\newcommand{\ket}[1]{\left\vert #1 \right\rangle }
\newcommand{\vev}[1]{\left\langle  #1 \right\rangle }
\newcommand{\ul}[1]{\underline{#1}}

\makeatother
%%%%%%%%% End of private macros %%%%%%%%%%%

%%%%%%%%%%%%%%%%%%%%%%%%%%%%%%%%%%%%%%%%%%%%%%%%%%%%%%%%%%%%%%%%%%%
\begin{titlepage}
\title{
\vspace{1cm}
First Order Symmetry Operators for the Linearized Field Equation of Metric Perturbations
}
\author{Yoji Michishita
\thanks{
{\tt michishita@edu.kagoshima-u.ac.jp}
}
\\[7pt]
{\it Department of Physics, Faculty of Education, Kagoshima University}\\
{\it Kagoshima, 890-0065, Japan}
}

\date{\normalsize September, 2019}
\maketitle
\thispagestyle{empty}

\begin{abstract}
\normalsize
We determine the general form of the first order linear symmetry operators for the linearized field equation of metric
perturbations in the spacetimes of dimension $D\ge 4$. 
Apart from the part derived easily from the invariance under general coordinate transformations, we find 
a part consisting of a Killing-Yano 3-form.
\end{abstract}
\end{titlepage}

\clearpage
%%%%%%%%%%%%%%%%%%%%%%%%%%%%%%%%%%%%%%%%%%%%%%%%%%%%%%%%%%%%%%%%%%%
\section{Introduction}

It is often necessary to solve the linearized equations of motion of various fields in given background geometries.
Especially the equation of the perturbations of the metric $h_{\mu\nu}$ is important:
Its time-dependent solutions indicate how gravitational waves propagate or whether the background is stable. 
It may also give moduli of the background geometry satisfying the background equation of motion.
To make it easier to solve the linearized equation $M_{\lambda\rho}{}^{\mu\nu}h_{\mu\nu}=0$,
where $M_{\lambda\rho}{}^{\mu\nu}$ is a second order derivative operator defined later in \eqref{leqop},
we consider symmetry operators of $M_{\lambda\rho}{}^{\mu\nu}$ i.e.
pairs of operators $(Q_{\lambda\rho}{}^{\mu\nu}, S_{\lambda\rho}{}^{\mu\nu})$ satisfying
\beq
Q_{\lambda\rho}{}^{\kappa\phi}M_{\kappa\phi}{}^{\mu\nu}h_{\mu\nu}
=M_{\lambda\rho}{}^{\kappa\phi}S_{\kappa\phi}{}^{\mu\nu}h_{\mu\nu}.
\eeq
Such operators, especially lower order ones, for lower spin fields, or
in some classes of backgrounds have been investigated in the literature.
Often variants of symmetry operators, such as the ones for the equations of lower spin fields obtained
from the original equation by some ansatz for the form of solutions, or the ones
connecting solutions to the original equation with solutions to easier equations, have also been considered. 
The following is a partial list of recent discussion:
For spin 0 fields in generalized Kerr-NUT-(A)dS spacetimes, see \cite{sk07}. 
For spin 1/2 fields see e.g. \cite{ms79,bc96,bk04,ckk11}.
For spin 1 fields under some ansatz in Kerr-NUT-(A)dS spacetimes see \cite{kfk18,hty19}, and
in 4 dimensions see \cite{abb14}.
For spin 3/2 fields see e.g. \cite{m18}.
For spin 2 and lower spin fields in Petrov type D spacetimes and higher dimensional extensions, see e.g. \cite{ab16,a16,a17}.

In this paper, we make no assumption about the form and the signature of the background geometry except that
the dimension $D$ is greater than or equal to 4, and determine the general form of the first order symmetry operators
$(Q_{\lambda\rho}{}^{\mu\nu}, S_{\lambda\rho}{}^{\mu\nu})$ for the equation of $h_{\mu\nu}$
by straightforward calculations. We do not use background equation of motion at intermediate steps,
and at the final step we use it and find the results \eqref{solQ} and \eqref{solS}.
Nontrivial parts of them consist of three parts: a gauge transformation part and a part consisting of a (conformal) Killing vector, 
which are well known and can be derived easily from the invariance under general coordinate transformations,
and a part consisting of a Killin-Yano 3-form.
In section 2 we give conditions for first order symmetry operators, and
in section 3 we show the general solution of the conditions and give conclusions for readers who are interested
only in the results. The details of how to solve the conditions is shown in section 4.
In Appendix we briefly summarize 
properties of conformal Killing vectors and Killing-Yano 3-forms used in the calculations.

%%%%%%%%%%%%%%%%%%%%%%%%%%%%%%%%%%%%%%%%%%%%%%%%%%%%%%%%%%%%%%%%%%%
\section{Preliminaries}

We consider the metric $g_{\mu\nu}$ in $D$-dimensional spaces, which we decompose into the background
metric $G_{\mu\nu}$ and the perturbation $h_{\mu\nu}$: $g_{\mu\nu}=G_{\mu\nu}+h_{\mu\nu}$.
No assumption about the signature of $g_{\mu\nu}$ is made.

Einstein-Hlibert Lagrangian for $g_{\mu\nu}$ (up to the overall constant factor) is given by 
\beq
\mathcal{L}=\sqrt{|g|}(R(g)-2\Lambda).
\eeq
The part linear in $h_{\mu\nu}$ is given by
\beq
\mathcal{L}\big|_h = 
 -\sqrt{|G|}\Big[R^{\mu\nu}(G)-\frac{1}{2}G^{\mu\nu}R(G)+\Lambda G^{\mu\nu}\Big]h_{\mu\nu}
 + \text{(total derivative term)},
\eeq
which shows that if there is no matter field the background equation of motion is
the following vacuum Einstein equation:
\beq
R^{\mu\nu}(G)-\frac{1}{2}G^{\mu\nu}R(G)+\Lambda G^{\mu\nu}=0,
\label{bveom0}
\eeq
or equivalently,
\beq
R^{\mu\nu}(G) = \frac{2}{D-2}\Lambda G^{\mu\nu},
\label{bveom1}
\eeq
and the following also holds:
\beq
R(G) = \frac{2D}{D-2}\Lambda.
\label{bveom2}
\eeq
Unless otherwise stated, we do not impose these equations on the background.
The part quadratic in $h_{\mu\nu}$ is given by
\beq
\mathcal{L}\big|_{h^2} = \sqrt{|G|}h_{\lambda\rho}M^{\lambda\rho\mu\nu}h_{\mu\nu}
 + \text{(total derivative term)},
\eeq
where $M^{\lambda\rho\mu\nu}$ is the following hermitian operator:
\bea
M_{\lambda\rho}{}^{\mu\nu} & = & \frac{1}{4}(
 \delta_\lambda{}^{(\mu}\delta_\rho{}^{\nu)}G^{\sigma\tau} - G_{\lambda\rho}G^{\mu\nu}G^{\sigma\tau}
 + \delta_{(\lambda}{}^\sigma\delta_{\rho)}{}^\tau G^{\mu\nu}
\nn & &
 + G_{\lambda\rho}G^{(\mu|\sigma}G^{\tau|\nu)}
 -2 \delta_{(\lambda}{}^\tau\delta_{\rho)}{}^{(\mu}G^{\nu)\sigma}
)\nabla_\sigma\nabla_\tau
\nn & &
 + \delta_{(\lambda}{}^{(\mu}R_{\rho)}{}^{\nu)}(G)
 + \frac{1}{8}G_{\lambda\rho}G^{\mu\nu}R(G)  - \frac{1}{4}\delta_{(\lambda}{}^\mu\delta_{\rho)}{}^\nu R(G)
\nn & &
 - \frac{1}{4}G_{\lambda\rho}R^{\mu\nu}(G) - \frac{1}{4}G^{\mu\nu}R_{\lambda\rho}(G)
\nn & &
-\frac{1}{4}\Lambda(G_{\lambda\rho}G^{\mu\nu}-2\delta_{(\lambda}{}^\mu\delta_{\rho)}{}^\nu),
\label{leqop}
\eea
where $\nabla_\mu$ is the covariant derivative of the background geometry, and 
indices are raised and lowered by $G^{\mu\nu}$ and $G_{\mu\nu}$.
Henceforth the background Riemann tensor $R_{\lambda\rho\mu\nu}(G)$ is denoted just by $R_{\lambda\rho\mu\nu}$.

The linearized equation of motion for the perturbation is $M_{\lambda\rho}{}^{\mu\nu}h_{\mu\nu}=0$.
If operators $Q_{\lambda\rho}{}^{\mu\nu}$ and $S_{\lambda\rho}{}^{\mu\nu}$ satisfying
\beq
Q_{\lambda\rho}{}^{\kappa\phi}M_{\kappa\phi}{}^{\mu\nu}h_{\mu\nu}
=M_{\lambda\rho}{}^{\kappa\phi}S_{\kappa\phi}{}^{\mu\nu}h_{\mu\nu}
\label{symopcond0}
\eeq
exist, then we can generate a new solution $S_{\lambda\rho}{}^{\mu\nu}h_{\mu\nu}$ from a solution $h_{\mu\nu}$.
If $S_{\lambda\rho}{}^{\mu\nu}$ is a commuting operator i.e. 
$Q_{\lambda\rho}{}^{\mu\nu}=S_{\lambda\rho}{}^{\mu\nu}$, we can use it to classify the solutions
by the simultaneous diagonalization with $M_{\lambda\rho}{}^{\mu\nu}$.
Such operators are called symmetry operators.
In this paper we determine the general form of symmetry operators
which are first order in derivative operator i.e. 
from the condition \eqref{symopcond0} for $Q_{\lambda\rho}{}^{\mu\nu}$ and $S_{\lambda\rho}{}^{\mu\nu}$
in the following form:
\bea
Q_{\lambda\rho}{}^{\mu\nu} & = &
 Q_{\lambda\rho}{}^{\mu\nu\phi}\nabla_\phi + q_{\lambda\rho}{}^{\mu\nu},
\\ 
S_{\lambda\rho}{}^{\mu\nu} & = &
 S_{\lambda\rho}{}^{\mu\nu\phi}\nabla_\phi + s_{\lambda\rho}{}^{\mu\nu},
\eea
we determine $Q_{\lambda\rho}{}^{\mu\nu\phi}$, $q_{\lambda\rho}{}^{\mu\nu}$,
$S_{\lambda\rho}{}^{\mu\nu\phi}$, and $s_{\lambda\rho}{}^{\mu\nu}$.

Some examples of the symmetry operators can be found easily.
The first example is a trivial one:
If $Q_{\lambda\rho}{}^{\mu\nu}$ and $S_{\lambda\rho}{}^{\mu\nu}$ are proportional to the unit matrix:
\beq
S_{\lambda\rho}{}^{\mu\nu} = Q_{\lambda\rho}{}^{\mu\nu}
 = c \delta_{(\lambda}{}^\mu\delta_{\rho)}{}^\nu,
\quad c= \text{const.}
\label{trivial0}
\eeq
\eqref{symopcond0} is satisfied.
To give next examples, we consider infinitesimal general coordinate transformation $x^{\prime\mu}=x^\mu-\epsilon^\mu$. 
$g_{\mu\nu}$ is transformed by this, which we regard as the transformation of the perturbation $h_{\mu\nu}$:
\beq
\delta g_{\mu\nu} = \delta(G_{\mu\nu}+h_{\mu\nu})=G_{\mu\nu}+\delta h_{\mu\nu},
\eeq
and
\beq
\delta h_{\mu\nu} = \nabla_\mu\epsilon_\nu + \nabla_\nu\epsilon_\mu + \mathcal{L}_\epsilon h_{\mu\nu},
\label{gct}
\eeq
where $\mathcal{L}_\epsilon$ is the Lie derivative operator along $\epsilon^\mu$:
\bea
\mathcal{L}_\epsilon h_{\mu\nu} & = &
 \epsilon^\lambda\p_\lambda h_{\mu\nu} 
 + \p_\mu\epsilon^\lambda h_{\lambda\nu} + \p_\nu\epsilon^\lambda h_{\mu\lambda}
\nn & = & 
 \epsilon^\lambda\nabla_\lambda h_{\mu\nu} 
 + \nabla_\mu\epsilon^\lambda h_{\lambda\nu} + \nabla_\nu\epsilon^\lambda h_{\mu\lambda}.
\eea
Under this transformation, $\delta\mathcal{L}=\p_\mu(\epsilon^\mu\mathcal{L})$ for arbitrary $h_{\mu\nu}$,
and therefore
\bea
0 & = & \int d^Dx \delta\mathcal{L}
\nn & = & 
 \int d^Dx
 \sqrt{|G|}\Big[-\Big(R^{\mu\nu}(G)-\frac{1}{2}G^{\mu\nu}R(G)+\Lambda G^{\mu\nu}\Big)\delta h_{\mu\nu}
\nn & & 
 +\delta h_{\lambda\rho}M^{\lambda\rho\mu\nu}h_{\mu\nu}
 +h_{\lambda\rho}M^{\lambda\rho\mu\nu}\delta h_{\mu\nu}
 +O(h^2, \delta h)
\Big].
\eea
If $\epsilon^\mu$ is a Killing vector of the background geometry,
$\delta h_{\mu\nu}$ contains only terms linear in $h_{\mu\nu}$, and 
from the terms quadratic in $h_{\mu\nu}$ in the above,
\bea
0 & = & \int d^Dx \sqrt{|G|}\Big[
 (\mathcal{L}_\epsilon h_{\lambda\rho})M^{\lambda\rho\mu\nu}h_{\mu\nu}
 +h_{\lambda\rho}M^{\lambda\rho\mu\nu}\mathcal{L}_\epsilon h_{\mu\nu}
\Big]
\nn & = & 
 \int d^Dx \sqrt{|G|}\Big[ h_{\lambda\rho}
 (M^{\lambda\rho\mu\nu}\mathcal{L}_\epsilon - \mathcal{L}_\epsilon M^{\lambda\rho\mu\nu})h_{\mu\nu}
\Big].
\eea
Therefore we expect that $\mathcal{L}_\epsilon$ commutes with $M_{\lambda\rho}{}^{\mu\nu}$:
\beq
\mathcal{L}_\epsilon M_{\lambda\rho}{}^{\mu\nu}h_{\mu\nu} = M_{\lambda\rho}{}^{\mu\nu}\mathcal{L}_\epsilon h_{\mu\nu}.
\label{symKV}
\eeq
Indeed this can be directly confirmed by the fact that $\mathcal{L}_\epsilon$ commutes with $\nabla_\mu$ and 
$\mathcal{L}_\epsilon R^\lambda{}_{\rho\mu\nu}=0$.
Note that \eqref{symKV} is true even when the background geometry does not satisfy the equation of motion.
If the background geometry satisfies the vacuum equation of motion, 
$\mathcal{L}$ has no linear term in $h_{\mu\nu}$, and
from the terms in $\delta\mathcal{L}$ linear in $h_{\mu\nu}$,
\beq
0 = \int d^Dx \sqrt{|G|}\Big[
 (\nabla^\lambda\epsilon^\rho) M_{\lambda\rho}{}^{\mu\nu}h_{\mu\nu}
 + h^{\lambda\rho}M_{\lambda\rho}{}^{\mu\nu}\nabla_\mu\epsilon_\nu
\Big].
\eeq
Since this is true for arbitrary $h_{\mu\nu}$ and $\epsilon^\mu$,
\beq
M_{\lambda\rho}{}^{\mu\nu}\nabla_\mu\epsilon_\nu=0,
\label{trivial1}
\eeq
\beq
\epsilon^\lambda\nabla^\rho M_{\lambda\rho}{}^{\mu\nu}=0.
\label{trivial2}
\eeq
These give eigenvectors of zero eigenvalue for $M_{\lambda\rho}{}^{\mu\nu}$ and are regarded
as the degrees of freedom of gauge transformation of $h_{\mu\nu}$.
We will see these examples appear in the general form of the first order symmetry operators.

Let us write down the condition \eqref{symopcond0} for each order of derivative.
Note that indices on covariant derivatives can be symmetrized by the followings:
\bea
\nabla_\sigma\nabla_\tau h_{\mu\nu}
 & = & \nabla_{(\sigma}\nabla_{\tau)}h_{\mu\nu} + R_{\sigma\tau(\mu}{}^\kappa h_{\nu)\kappa},
\\
\nabla_\phi\nabla_\sigma\nabla_\tau h_{\mu\nu}
& = & \nabla_{(\phi}\nabla_\sigma\nabla_{\tau)}h_{\mu\nu}
 + R_{\phi(\sigma|\mu|}{}^\kappa\nabla_{\tau)}h_{\kappa\nu}
 + R_{\phi(\sigma|\nu|}{}^\kappa\nabla_{\tau)}h_{\mu\kappa}
\nn & &
 + \frac{1}{2}R_{\sigma\tau\mu}{}^\kappa\nabla_\phi h_{\kappa\nu}
 + \frac{1}{2}R_{\sigma\tau\nu}{}^\kappa\nabla_\phi h_{\mu\kappa}
 + \frac{2}{3}R_{\phi(\sigma\tau)}{}^\kappa\nabla_\kappa h_{\mu\nu}
\nn & &
 + \frac{2}{3}\nabla_{(\sigma}R_{\phi)\tau\mu}{}^\kappa h_{\kappa\nu}
 + \frac{2}{3}\nabla_{(\sigma}R_{\phi)\tau\nu}{}^\kappa h_{\mu\kappa}.
\eea
Then terms proportional to $\nabla_{(\phi}\nabla_\sigma\nabla_{\tau)}h_{\mu\nu}$, 
$\nabla_{(\phi}\nabla_{\tau)}h_{\mu\nu}$, $\nabla_{\phi}h_{\mu\nu}$, and $h_{\mu\nu}$ in \eqref{symopcond0}
cancel separately.
From the part proportional to $\nabla_{(\phi}\nabla_\sigma\nabla_{\tau)}h_{\mu\nu}$,
\bea
\lefteqn{ Q_{\lambda\rho\mu\nu}{}^{(\phi}G^{\sigma\tau)} - Q_{\lambda\rho\kappa}{}^{\kappa(\phi}G^{\sigma\tau)}G_{\mu\nu}
 + Q_{\lambda\rho}{}^{(\sigma\tau\phi)}G_{\mu\nu}
} \nn 
\lefteqn{ + Q_{\lambda\rho\kappa}{}^{\kappa(\phi}\delta_\mu{}^\sigma\delta_\nu{}^{\tau)}
 -2 Q_{\lambda\rho}{}^{(\tau}{}_{(\mu}{}^{\phi}\delta_{\nu)}{}^{\sigma)}
} \nn & = &
 S_{\lambda\rho\mu\nu}{}^{(\phi}G^{\sigma\tau)} - S_\kappa{}^\kappa{}_{\mu\nu}{}^{(\phi}G^{\sigma\tau)}G_{\lambda\rho}
 + S_\kappa{}^\kappa{}_{\mu\nu}{}^{(\phi}\delta_\lambda{}^\sigma\delta_\rho{}^{\tau)}
\nn & &
 + S^{(\sigma\tau}{}_{\mu\nu}{}^{\phi)}G_{\lambda\rho}
 -2 S_{(\lambda}{}^{(\sigma}{}_{|\mu\nu|}{}^\phi\delta_{\rho)}{}^{\tau)}.
\label{Ho3}
\eea
From the part proportional to $\nabla_{(\phi}\nabla_{\tau)}h_{\mu\nu}$,
\bea
\lefteqn{ q_{\lambda\rho}{}^{\mu\nu}G^{\tau\phi} - q_{\lambda\rho\kappa}{}^\kappa G^{\tau\phi}G^{\mu\nu}
 + q_{\lambda\rho}{}^{\tau\phi}G^{\mu\nu} + q_{\lambda\rho\kappa}{}^\kappa G^{\tau(\mu}G^{\nu)\phi}
 - q_{\lambda\rho}{}^{\tau(\mu}G^{\nu)\phi} - q_{\lambda\rho}{}^{\phi(\mu}G^{\nu)\tau}
} \nn & = & 
 s_{\lambda\rho}{}^{\mu\nu}G^{\tau\phi} - s_\kappa{}^{\kappa\mu\nu}G_{\lambda\rho}G^{\tau\phi}
 + s_\kappa{}^{\kappa\mu\nu}\delta_{(\lambda}{}^\tau\delta_{\rho)}{}^\phi
 + s^{\tau\phi\mu\nu}G_{\lambda\rho} - 2s_{(\lambda}{}^{(\tau|\mu\nu|}\delta_{\rho)}{}^{\phi)}
\nn & &
 +2\nabla^{(\tau}S_{\lambda\rho}{}^{|\mu\nu|\phi)}
 -2\nabla^{(\tau}S_\kappa{}^{|\kappa\mu\nu|\phi)}G_{\lambda\rho}
 +2\nabla_\sigma S^\sigma{}^{(\tau|\mu\nu|\phi)}G_{\lambda\rho}
\nn & &
 +2\nabla_{(\lambda}S_{|\kappa|}{}^{\kappa\mu\nu(\tau}\delta^{\phi)}{}_{\rho)}
 -2\nabla_{(\lambda}S_{\rho)}{}^{(\tau|\mu\nu|\phi)}
 -2\nabla_\kappa S_{(\lambda}{}^{\kappa\mu\nu(\tau}\delta_{\rho)}{}^{\phi)}.
\label{ho2}
\eea
From the part proportional to $\nabla_{\phi}h_{\mu\nu}$,
\bea
\lefteqn{ Q_{\lambda\rho}{}^{(\mu|\kappa\sigma}R_\sigma{}^{\phi|}{}_\kappa{}^{\nu)}
 -\frac{1}{2}Q_{\lambda\rho\kappa}{}^{\kappa\sigma}R_\sigma{}^{(\mu}G^{\nu)\phi}
 -\frac{1}{6}Q_{\lambda\rho\kappa}{}^{\kappa\sigma}R_\sigma{}^{(\mu\nu)\phi}
 +\frac{1}{2}Q_{\lambda\rho}{}^{\phi\kappa\sigma}R_\sigma{}^{(\mu\nu)}{}_\kappa
} \nn \lefteqn{
 +\frac{1}{2}Q_{\lambda\rho}{}^{\phi(\mu|\kappa|}R_\kappa{}^{\nu)}
 -\frac{1}{2}Q_{\lambda\rho}{}^{\sigma\tau\kappa}R_{\kappa\sigma\tau}{}^{(\mu}G^{\nu)\phi}
 +\frac{1}{6}Q_{\lambda\rho}{}^{\kappa(\mu|\sigma|}R_{\sigma\kappa}{}^{\nu)\phi}
 -\frac{1}{2}Q_{\lambda\rho}{}^{\kappa\sigma\phi}R_\kappa{}^{(\mu\nu)}{}_\sigma
} \nn \lefteqn{
 +\frac{3}{2}Q_{\lambda\rho}{}^{\kappa(\mu|\phi|}R_\kappa{}^{\nu)}
 -\frac{1}{3}Q_{\lambda\rho}{}^{\mu\nu\kappa}R_\kappa{}^\phi
 +\frac{1}{3}Q_{\lambda\rho\kappa}{}^{\kappa\sigma}R_\sigma{}^\phi G^{\mu\nu}
 +\frac{1}{3}Q_{\lambda\rho}{}^{\sigma\tau\kappa}R_{\kappa\sigma\tau}{}^\phi G^{\mu\nu}
} \nn \lefteqn{
 -\frac{1}{3}Q_{\lambda\rho}{}^{\kappa(\mu|\sigma|}R_\sigma{}^{\nu)}{}_\kappa{}^\phi
 +\frac{1}{4}Q_{\lambda\rho\kappa}{}^{\kappa\phi}R G^{\mu\nu}
 -\frac{1}{2}Q_{\lambda\rho}{}^{\mu\nu\phi}R
 -\frac{1}{2}Q_{\lambda\rho\kappa}{}^{\kappa\phi}R^{\mu\nu}
 -\frac{1}{2}Q_{\lambda\rho}{}^{\sigma\tau\phi}R_{\sigma\tau}G^{\mu\nu}
} \nn \lefteqn{
 -\frac{\Lambda}{2}(Q_{\lambda\rho\kappa}{}^{\kappa\phi}G^{\mu\nu} - 2Q_{\lambda\rho}{}^{\mu\nu\phi})
} \nn & = & 
 \frac{1}{2}\nabla_\kappa\nabla^\kappa S_{\lambda\rho}{}^{\mu\nu\phi}
 -\frac{1}{2}G_{\lambda\rho}\nabla_\sigma\nabla^\sigma S_\kappa{}^{\kappa\mu\nu\phi}
 +\frac{1}{2}\nabla_{(\lambda}\nabla_{\rho)}S_\kappa{}^{\kappa\mu\nu\phi}
\nn & & 
 +\frac{1}{2}G_{\lambda\rho}\nabla^\sigma\nabla^\tau S_{\sigma\tau}{}^{\mu\nu\phi}
 -\nabla_\kappa\nabla_{(\lambda}S_{\rho)}{}^{\kappa\mu\nu\phi}
\nn & & 
 +2S_{(\lambda}{}^{\kappa\mu\nu\phi}R_{\rho)\kappa}
 +\frac{1}{4}S_\kappa{}^{\kappa\mu\nu\phi}RG_{\lambda\rho}
 -\frac{1}{2}S_{\lambda\rho}{}^{\mu\nu\phi}R
 -\frac{1}{2}S_{\sigma\tau}{}^{\mu\nu\phi}R^{\sigma\tau}G_{\lambda\rho} 
\nn & & 
 -\frac{1}{2}S_\kappa{}^{\kappa\mu\nu\phi}R_{\lambda\rho}
 -S_{\lambda\rho}{}^{\sigma(\mu|\kappa}R_\kappa{}^{\phi|}{}_\sigma{}^{\nu)}
 +\frac{1}{6}S_{\lambda\rho}{}^{\mu\nu\kappa}R_\kappa{}^\phi
 +S_\kappa{}^{\kappa\sigma(\mu|\tau}R_\tau{}^{\phi|}{}_\sigma{}^{\nu)}G_{\lambda\rho}
\nn & & 
 -\frac{1}{6}S_\kappa{}^{\kappa\mu\nu\sigma}R_\sigma{}^\phi G_{\lambda\rho}
 -S_\kappa{}^{\kappa\sigma(\mu|\tau|}R_{\tau(\lambda|\sigma|}{}^{\nu)}\delta_{\rho)}{}^\phi
 -\frac{1}{6}S_\kappa{}^{\kappa\mu\nu\sigma}R_{\sigma(\lambda\rho)}{}^\phi
 -S^{\kappa\phi\sigma(\mu|\tau|}R_{\tau\kappa\sigma}{}^{\nu)}G_{\lambda\rho}
\nn & & 
 -\frac{1}{6}S_{\sigma\tau}{}^{\mu\nu\kappa}R_\kappa{}^{\sigma\tau\phi}G_{\lambda\rho}
 +S_{(\lambda}{}^{\kappa\sigma(\mu|\phi|}R_{\rho)\kappa\sigma}{}^{\nu)}
 -S_{(\lambda}{}^{\kappa\sigma(\mu|\tau|}R_{|\kappa\tau\sigma|}{}^{\nu)}\delta_{\rho)}{}^\phi
 -S_{(\lambda}{}^{\phi\sigma(\mu|\tau|}R_{\rho)\tau\sigma}{}^{\nu)}
\nn & & 
 +\frac{1}{3}S_{(\lambda}{}^{\kappa\mu\nu\sigma}R_{\rho)\kappa\sigma}{}^\phi
 -\frac{1}{3}S_{(\lambda}{}^{\kappa\mu\nu\sigma}R_{|\kappa\sigma|\rho)}{}^\phi
\nn & & 
 +\nabla^\phi s_{\lambda\rho}{}^{\mu\nu}
 -G_{\lambda\rho}\nabla^\phi s_\kappa{}^{\kappa\mu\nu}
 +\delta_{(\lambda}{}^\phi\nabla_{\rho)}s_\kappa{}^{\kappa\mu\nu}
\nn & & 
 +G_{\lambda\rho}\nabla_\kappa s^{\kappa\phi\mu\nu}
 -\nabla_{(\lambda}s_{\rho)}{}^{\phi\mu\nu}
 -\delta_{(\lambda}{}^\phi\nabla_{|\kappa|}s_{\rho)}{}^{\kappa\mu\nu}
\nn & & 
 -\frac{\Lambda}{2}(S_\kappa{}^{\kappa\mu\nu\phi}G_{\lambda\rho} - 2S_{\lambda\rho}{}^{\mu\nu\phi}).
\label{Ho1}
\eea
From the part proportional to $h_{\mu\nu}$,
\bea
\lefteqn{ \frac{2}{3}Q_{\lambda\rho}{}^{(\mu|\kappa\sigma|}\nabla_\sigma R_\kappa{}^{\nu)}
 + \frac{1}{6}Q_{\lambda\rho}{}^{(\mu|\kappa\sigma|}\nabla^{\nu)}R_{\kappa\sigma}
 - \frac{1}{6}Q_{\lambda\rho\kappa}{}^{\kappa\sigma}\nabla_\sigma R^{\mu\nu}
 - \frac{1}{6}Q_{\lambda\rho\kappa}{}^{\kappa\sigma}\nabla^{(\mu}R_\sigma{}^{\nu)}
} \nn \lefteqn{ 
 - \frac{1}{6}Q_{\lambda\rho}{}^{\kappa\sigma\tau}\nabla^{(\mu}R_{\tau\kappa\sigma}{}^{\nu)}
 - \frac{1}{6}Q_{\lambda\rho}{}^{\kappa\sigma\tau}\nabla_\tau R^{(\mu}{}_{\kappa\sigma}{}^{\nu)}
 + \frac{1}{8}Q_{\lambda\rho\kappa}{}^{\kappa\sigma}G^{\mu\nu}\nabla_\sigma R
 - \frac{1}{4}Q_{\lambda\rho}{}^{\mu\nu\kappa}\nabla_\kappa R
} \nn \lefteqn{ 
 - \frac{1}{4}Q_{\lambda\rho}{}^{\sigma\tau\kappa}G^{\mu\nu}\nabla_\kappa R_{\sigma\tau}
} \nn \lefteqn{ 
 - \frac{1}{4}q_{\lambda\rho}{}^{\kappa\sigma}R^{(\mu}{}_{\kappa\sigma}{}^{\nu)}
 + \frac{3}{4}q_{\lambda\rho}{}^{\kappa(\mu}R_\kappa{}^{\nu)}
 - \frac{1}{4}q_{\lambda\rho\kappa}{}^\kappa R^{\mu\nu}
 - \frac{1}{4}q_{\lambda\rho}{}^{\sigma\tau}G^{\mu\nu}R_{\sigma\tau}
} \nn \lefteqn{ 
 - \frac{1}{4}q_{\lambda\rho}{}^{\mu\nu}R
 + \frac{1}{8}q_{\lambda\rho\kappa}{}^\kappa G^{\mu\nu}R
} \nn \lefteqn{ 
 -\frac{\Lambda}{4}(q_{\lambda\rho\kappa}{}^\kappa G^{\mu\nu} - 2q_{\lambda\rho}{}^{\mu\nu})
} \nn & = & 
 \frac{1}{3}S_{\lambda\rho}{}^{\kappa(\mu|\sigma|}\nabla_\kappa R^{\nu)}{}_\sigma
 - \frac{1}{3}S_{\lambda\rho}{}^{\kappa(\mu|\sigma|}\nabla^{\nu)}R_{\kappa\sigma}
 - \frac{1}{3}G_{\lambda\rho}S_\kappa{}^{\kappa\sigma(\mu|\tau|}\nabla_\sigma R^{\nu)}{}_\tau
\nn & & 
 + \frac{1}{3}G_{\lambda\rho}S_\kappa{}^{\kappa\sigma(\mu|\tau|}\nabla^{\nu)} R_{\sigma\tau}
 + \frac{1}{3}G_{\lambda\rho}S^{\sigma\tau\phi(\mu|\kappa|}\nabla_\sigma R_{\tau\kappa\phi}{}^{\nu)}
 + \frac{1}{3}S_\kappa{}^{\kappa\sigma(\mu|\tau|}\nabla_{(\lambda}R_{\rho)\tau\sigma}{}^{\nu)}
\nn & & 
 - \frac{1}{3}S_{(\lambda}{}^{\kappa\sigma(\mu|\tau|}\nabla_{\rho)}R_{\kappa\tau\sigma}{}^{\nu)}
 - \frac{1}{3}S_{(\lambda}{}^{\kappa\sigma(\mu|\tau|}\nabla_{|\kappa|}R_{\rho)\tau\sigma}{}^{\nu)}
 + \frac{1}{2}\nabla^\kappa S_{\lambda\rho}{}^{\sigma(\mu|\tau|}R_{\kappa\tau\sigma}{}^{\nu)}
\nn & & 
 - \frac{1}{2}\nabla^\sigma S_\kappa{}^{\kappa\tau(\mu|\phi|}G_{\lambda\rho}R_{\sigma\phi\tau}{}^{\nu)}
 + \frac{1}{2}\nabla_{(\lambda}S_{|\kappa|}{}^{\kappa\sigma(\mu|\tau|}R_{\rho)\tau\sigma}{}^{\nu)}
 + \frac{1}{2}\nabla_\sigma S^{\sigma\tau\kappa(\mu|\phi|}G_{\lambda\rho}R_{\tau\phi\kappa}{}^{\nu)}
\nn & & 
 - \frac{1}{2}\nabla_{(\lambda}S_{\rho)}{}^{\kappa\sigma(\mu|\tau|}R_{\kappa\tau\sigma}{}^{\nu)}
 - \frac{1}{2}\nabla_\kappa S_{(\lambda}{}^{\kappa\sigma(\mu|\tau|}R_{\rho)\tau\sigma}{}^{\nu)}
\nn & & 
 + \frac{1}{4}\nabla^\kappa\nabla_\kappa s_{\lambda\rho}{}^{\mu\nu}
 - \frac{1}{4}G_{\lambda\rho}\nabla^\sigma\nabla_\sigma s_\kappa{}^{\kappa\mu\nu}
 + \frac{1}{4}\nabla_{(\lambda}\nabla_{\rho)}s_\kappa{}^{\kappa\mu\nu}
\nn & & 
 + \frac{1}{4}G_{\lambda\rho}\nabla^\sigma\nabla^\tau s_{\sigma\tau}{}^{\mu\nu}
 - \frac{1}{2}\nabla_\kappa\nabla_{(\lambda}s_{\rho)}{}^{\kappa\mu\nu}
\nn & & 
 - \frac{1}{2}R_{\kappa(\lambda}{}^{\sigma(\mu} s_{\rho)}{}^{|\kappa|}{}_\sigma{}^{\nu)}
 + R_{(\lambda}{}^\kappa s_{\rho)\kappa}{}^{\mu\nu}
 - \frac{1}{4}R_{\lambda\rho}s_\kappa{}^{\kappa\mu\nu}
 - \frac{1}{4}G_{\lambda\rho}R^{\sigma\tau}s_{\sigma\tau}{}^{\mu\nu}
\nn & & 
 - \frac{1}{4}Rs_{\lambda\rho}{}^{\mu\nu}
 + \frac{1}{8}G_{\lambda\rho}Rs_\kappa{}^{\kappa\mu\nu}
\nn & & 
 -\frac{\Lambda}{4}(s_\kappa{}^{\kappa\mu\nu}G_{\lambda\rho} - 2s_{\lambda\rho}{}^{\mu\nu}).
\label{ho0}
\eea
In the following sections we solve these conditions. However in 2 dimensions, 
$R^{\lambda\rho}{}_{\mu\nu}=R\delta^{[\lambda}{}_\mu\delta^{\rho]}{}_\nu$,
and any manifold satisfies vacuum Einstein equation with no cosmological constant.
Indeed it can be shown that $M_{\lambda\rho}{}^{\mu\nu}h_{\mu\nu}=0$ holds for any $h_{\mu\nu}$ if $\Lambda=0$.
In 3 dimensions, independent components of Riemann tensor are given in terms of Ricci tensor: 
\beq
R^{\lambda\rho}{}_{\mu\nu} = -6 \delta^{[\lambda}{}_\mu\delta^\rho{}_\nu\delta^{\sigma]}{}_\tau
 \Big(R^\tau{}_\sigma-\frac{1}{2}\delta^\tau{}_\sigma R\Big),
\eeq
and this means that any background satisfying vacuum Einstein equation is a space of constant curvature, 
which has been well studied.
Therefore in the following sections we only consider cases of $D\ge 4$.

%%%%%%%%%%%%%%%%%%%%%%%%%%%%%%%%%%%%%%%%%%%%%%%%%%%%%%%%%%%%%%%%%%%
\section{Results and conclusions}

In this section we summarize our results and give conclusions so that readers who are not interested in the details of 
the procedure for solving the equations \eqref{Ho3}, \eqref{ho2}, \eqref{Ho1}, and \eqref{ho0} can skip them.
The details will be explained in the next section.

First we show the solution to equation \eqref{Ho3} which is purely algebraic:
\bea
Q_{\lambda\rho}{}^{\mu\nu\phi} & = & 
 \delta_{(\lambda}{}^\mu\delta_{\rho)}{}^\nu K^\phi
 + \delta_{(\lambda}{}^{(\mu}Y_{\rho)}{}^{\nu)}{}^\phi
 + G^{\phi(\mu}F_{\lambda\rho}{}^{\nu)},
\label{solH3-1}
\\
S_{\lambda\rho}{}^{\mu\nu\phi} & = & 
 \delta_{(\lambda}{}^\mu\delta_{\rho)}{}^\nu K^\phi
 + \delta_{(\lambda}{}^{(\mu}Y_{\rho)}{}^{\nu)}{}^\phi
 + \delta^\phi{}_{(\lambda}H^{\mu\nu}{}_{\rho)},
\label{solH3-2}
\eea
where $K^\mu$ is an arbitrary vector, and $Y_{\mu\nu\lambda}$ is an arbitrary antisymmetric tensor.
$F_{\lambda\rho}{}^\phi$ and $H_{\lambda\rho}{}^\phi$ are arbitrary tensors satisfying
\beq
F_{\lambda\rho}{}^\phi = F_{\rho\lambda}{}^\phi, \quad
H_{\lambda\rho}{}^\phi = H_{\rho\lambda}{}^\phi.
\eeq

Using the above we obtain the following solution to \eqref{ho2}:
\bea
q_{\lambda\rho}{}^{\mu\nu} & = & 
 \delta_{(\lambda}{}^{(\mu}(\nabla_{\rho)}K^{\nu)} - \nabla^{\nu)}K_{\rho)})
 + \delta_{(\lambda}{}^{\mu}\delta_{\rho)}{}^{\nu}\wt{c},
\label{solh2-1}
\\
s_{\lambda\rho}{}^{\mu\nu} & = & 
 \delta_{(\lambda}{}^{(\mu}(\nabla_{\rho)}K^{\nu)} - \nabla^{\nu)}K_{\rho)})
 + \delta_{(\lambda}{}^{\mu}\delta_{\rho)}{}^{\nu}\wt{c}
 -\frac{2}{D}\delta_{(\lambda}{}^{\mu}\delta_{\rho)}{}^{\nu}\nabla_\kappa K^\kappa
 + \nabla_{(\lambda}H^{\mu\nu}{}_{\rho)},
\label{solh2-2}
\eea
where $\wt{c}$ is a scalar function,
and it turns out that $K^\mu$ must be a conformal Killing vector, and 
$Y_{\mu\nu\lambda}$ must be a Killing-Yano tensor.

Then we find that \eqref{Ho1} is equivalent to the following three relations:
\bea
\p_\mu\wt{c} & = & 
 \frac{1}{D-1}\Big(R-\frac{2D}{D-2}\Lambda\Big)\Big(
  F_{\mu\kappa}{}^\kappa - \frac{1}{D-2}H_\kappa{}^\kappa{}_\mu\Big)
\nn & & +
 \frac{2}{(D-1)(D-2)}\Big(R^{\sigma\tau}-\frac{2}{D-2}\Lambda G^{\sigma\tau}\Big)
  H_{\sigma\tau\mu},
\label{solH1-1}
\\
\p_\mu(\nabla_\kappa K^\kappa) & = & 
 \frac{D}{(D-1)(D+2)}\Big(R-\frac{2D}{D-2}\Lambda\Big)
\nn & & \times
\Big(
  F_{\mu\kappa}{}^\kappa + \frac{1}{2}F_\kappa{}^\kappa{}_\mu
  + \frac{D-4}{2(D-2)}H_{\mu\kappa}{}^\kappa
  - \frac{1}{D-2}H_\kappa{}^\kappa{}_\mu\Big)
\nn & & +
 \frac{2D}{(D-1)(D-2)(D+2)}\Big(R^{\sigma\tau}-\frac{2}{D-2}\Lambda G^{\sigma\tau}\Big)
 \Big( H_{\mu\sigma\tau} + H_{\sigma\tau\mu}\Big),
\label{solH1-2}
\eea
\bea
\lefteqn{
\Big(R^{\mu\nu}-\frac{1}{2}RG^{\mu\nu}+\Lambda G^{\mu\nu}\Big)F_{\lambda\rho}{}^\phi 
} \nn & = &
 \frac{1}{2(D-1)(D+2)}\Big(R-\frac{2D}{D-2}\Lambda\Big)\Big(
 4G_{\lambda\rho}G^{\mu\nu}F^{\phi\kappa}{}_\kappa
 -4\delta_{(\lambda}{}^{(\mu}\delta_{\rho)}{}^{\nu)}F^{\phi\kappa}{}_\kappa
\nn & &
 -2DG^{\mu\nu}\delta_{(\lambda}{}^\phi F_{\rho)\kappa}{}^\kappa
 -2DG_{\lambda\rho}G^{\phi(\mu}F^{\nu)\kappa}{}_\kappa
 +2D\delta_{(\lambda}{}^\phi\delta_{\rho)}{}^{(\mu}F^{\nu)\kappa}{}_\kappa
 +2DG^{\phi(\mu}\delta_{(\lambda}{}^{\nu)}F_{\rho)\kappa}{}^\kappa
\nn & &
 -DG_{\lambda\rho}G^{\mu\nu}F_\kappa{}^{\kappa\phi}
 +D\delta_{(\lambda}{}^{(\mu}\delta_{\rho)}{}^{\nu)}F_\kappa{}^{\kappa\phi}
\nn & &
 +2G^{\mu\nu}\delta_{(\lambda}{}^\phi F^\kappa{}_{|\kappa|\rho)}
 +2G_{\lambda\rho}G^{\phi(\mu}F_\kappa{}^{|\kappa|\nu)}
 -2\delta_{(\lambda}{}^\phi\delta_{\rho)}{}^{(\mu}F_\kappa{}^{|\kappa|\nu)}
 -2G^{\phi(\mu}\delta_{(\lambda}{}^{\nu)}F^\kappa{}_{|\kappa|\rho)}
\nn & &
 -DG_{\lambda\rho}G^{\mu\nu}H^{\phi\kappa}{}_\kappa
 +D\delta_{(\lambda}{}^{(\mu}\delta_{\rho)}{}^{\nu)}H^{\phi\kappa}{}_\kappa
\nn & &
 +2G^{\mu\nu}\delta_{(\lambda}{}^\phi H_{\rho)\kappa}{}^\kappa
 +2G_{\lambda\rho}G^{\phi(\mu}H^{\nu)\kappa}{}_\kappa
 -2\delta_{(\lambda}{}^\phi\delta_{\rho)}{}^{(\mu}H^{\nu)\kappa}{}_\kappa
 -2G^{\phi(\mu}\delta_{(\lambda}{}^{\nu)}H_{\rho)\kappa}{}^\kappa
 \Big)
\nn & &
 + \frac{2}{(D-1)(D-2)(D+2)}\Big(R^{\sigma\tau}-\frac{1}{2}RG^{\sigma\tau}+\Lambda G^{\sigma\tau}\Big)
\nn & &
  \times\Big( -DG_{\lambda\rho}G^{\mu\nu}H^\phi{}_{\sigma\tau}
   + D\delta_{(\lambda}{}^{(\mu}\delta_{\rho)}{}^{\nu)}H^\phi{}_{\sigma\tau}
\nn & & 
   + 2\delta_{(\lambda}{}^\phi G^{\mu\nu}H_{\rho)\sigma\tau}
   + 2G_{\lambda\rho}G^{\phi(\mu}H^{\nu)}{}_{\sigma\tau}
   - 2\delta_{(\lambda}{}^\phi\delta_{\rho)}{}^{(\mu}H^{\nu)}{}_{\sigma\tau}
   - 2G^{\phi(\mu}\delta_{(\lambda}{}^{\nu)}H_{\rho)\sigma\tau}
\nn & & 
  + 2G_{\lambda\rho}G^{\mu\nu}H_{\sigma\tau}{}^\phi
  - 2\delta_{(\lambda}{}^{(\mu}\delta_{\rho)}{}^{\nu)}H_{\sigma\tau}{}^\phi
\nn & & 
  - D\delta_{(\lambda}{}^\phi G^{\mu\nu}H_{|\sigma\tau|\rho)}
  - DG_{\lambda\rho}G^{\phi(\mu}H_{\sigma\tau}{}^{\nu)}
  + D\delta_{(\lambda}{}^\phi\delta_{\rho)}{}^{(\mu}H_{\sigma\tau}{}^{\nu)}
  + DG^{\phi(\mu}\delta_{(\lambda}{}^{\nu)}H_{|\sigma\tau|\rho)}
\Big)
\nn & &
 + 2\Big(R_{(\lambda}{}^{(\mu}-\frac{1}{2}R\delta_{(\lambda}{}^{(\mu}+\Lambda\delta_{(\lambda}{}^{(\mu}
 \Big)H^{\nu)\phi}{}_{\rho)}
 - \Big(R_{\lambda\rho}-\frac{1}{2}RG_{\lambda\rho}+\Lambda G_{\lambda\rho}\Big)H^{\phi(\mu\nu)}
\nn & &
 + G_{\lambda\rho}\Big(R^{\sigma(\mu}-\frac{2}{D-2}\Lambda G^{\sigma(\mu}\Big)Y^{\nu)\phi}{}_\sigma
 + G^{\mu\nu}\Big(R^\sigma{}_{(\lambda}-\frac{2}{D-2}\Lambda\delta^\sigma{}_{(\lambda}\Big)Y_{\rho)}{}^\phi{}_\sigma
\nn & &
 - \delta_{(\lambda}{}^\phi\Big(R^{\sigma(\mu}-\frac{2}{D-2}\Lambda G^{\sigma(\mu}\Big)Y^{\nu)}{}_{\rho)\sigma}
 + G^{\phi(\mu}\Big(R^{|\sigma|}{}_{(\lambda}-\frac{2}{D-2}\Lambda G^{|\sigma|}{}_{(\lambda}\Big)Y^{\nu)}{}_{\rho)\sigma}
\nn & &
 + \delta_{(\lambda}{}^{(\mu}\Big(R^{\nu)\sigma}-\frac{2}{D-2}\Lambda G^{\nu)\sigma}\Big)Y^\phi{}_{\rho)\sigma}
 - \delta_{(\lambda}{}^{(\mu}\Big(R_{\rho)}{}^{|\sigma|}-\frac{2}{D-2}\Lambda G_{\rho)}{}^{|\sigma|}\Big)Y^{\nu)\phi}{}_\sigma.
\label{solH1-3}
\eea
Note that in these equations Riemann tensor (with no indices contracted) does not appear, and
Ricci tensor and $\Lambda$ appear only in the form of the background equations of motion
\eqref{bveom0}, \eqref{bveom1}, or \eqref{bveom2}.
The last equation \eqref{solH1-3} is somewhat complicated and we obtain no simpler relation from it.
However, as is clear from its expression, it is trivially satisfied if \eqref{bveom0} or \eqref{bveom1} is satisfied,
and \eqref{solH1-1} and \eqref{solH1-2} mean that $\wt{c}$ and $\nabla_\kappa K^\kappa$ are constants. 

Finally we find that \eqref{ho0} is equivalent to the following three relations:
\bea
0 & = &
\frac{(D-2)}{8D^2(D-1)}(G_{\lambda\rho}G^{\mu\nu} - D\delta_{(\lambda}{}^\mu \delta_{\rho)}{}^\nu)
\nabla_\kappa\Big[\Big(R-\frac{2D}{D-2}\Lambda\Big) F^{\kappa\sigma}{}_\sigma\Big]
\nn & &
 + \frac{(D-4)}{4(D-2)}\Big(R-\frac{2D}{D-2}\Lambda\Big)\delta_{(\lambda}{}^{(\mu}\nabla_{|\kappa|}H_{\rho)}{}^{\nu)\kappa}
\nn & &
 + \frac{1}{4D(D-2)}\Big(R-\frac{2D}{D-2}\Lambda\Big)
  \Big(DG_{\lambda\rho}\nabla_\kappa H^{\mu\nu\kappa} - (D-4)G^{\mu\nu}\nabla_\kappa H_{\lambda\rho}{}^\kappa\Big)
\nn & &
 -\frac{1}{8D(D-1)(D-2)}\Big(R-\frac{2D}{D-2}\Lambda\Big)
  \Big[(D+2)G_{\lambda\rho}G^{\mu\nu}
  + D(D-4)\delta_{(\lambda}{}^\mu\delta_{\rho)}{}^\nu\Big]\nabla_\kappa H_\sigma{}^{\sigma\kappa}
\nn & &
 +\frac{1}{4D^2(D-1)}\Big(R-\frac{2D}{D-2}\Lambda\Big)
  (D\delta_{(\lambda}{}^\mu\delta_{\rho)}{}^\nu - G_{\lambda\rho}G^{\mu\nu})\nabla_\kappa H^{\kappa\sigma}{}_\sigma
\nn & &
 +\frac{1}{8D(D-1)}\p_\kappa\Big(R-\frac{2D}{D-2}\Lambda\Big)
 (G_{\lambda\rho}G^{\mu\nu} - D\delta_{(\lambda}{}^\mu\delta_{\rho)}{}^\nu)H_\sigma{}^{\sigma\kappa}
\nn & &
 +\frac{1}{4D}\Big(R-\frac{2D}{D-2}\Lambda\Big)G^{\mu\nu}\nabla_{(\lambda}H^\kappa{}_{|\kappa|\rho)}
\nn & &
 +\frac{1}{4D}\p_\kappa\Big(R-\frac{2D}{D-2}\Lambda\Big)
 (D\delta_{(\lambda}{}^{(\mu}H_{\rho)}{}^{\nu)\kappa} - G^{\mu\nu}H_{\lambda\rho}{}^\kappa)
\nn & &
 -\frac{1}{4}\Big(R-\frac{2D}{D-2}\Lambda\Big)\nabla_{(\lambda}H^{\mu\nu}{}_{\rho)}
\nn & &
 + \frac{1}{2D^2(D-1)}\Big(R_{\kappa\sigma}-\frac{2}{D-2}\Lambda G_{\kappa\sigma}\Big)
  (G_{\lambda\rho}G^{\mu\nu} - D\delta_{(\lambda}{}^\mu\delta_{\rho)}{}^\nu)\nabla^\kappa H^{\sigma\tau}{}_\tau
\nn & &
 +\frac{1}{2(D-1)(D-2)}\Big(R_{\kappa\sigma}-\frac{2}{D-2}\Lambda G_{\kappa\sigma}\Big)
  (G_{\lambda\rho}G^{\mu\nu} - \delta_{(\lambda}{}^\mu\delta_{\rho)}{}^\nu)\nabla^\kappa H_\tau{}^{\tau\sigma}
\nn & &
 +\frac{1}{D-2}\Big(R_{\kappa\sigma}-\frac{2}{D-2}\Lambda G_{\kappa\sigma}\Big)
  \delta_{(\lambda}{}^{(\mu}\nabla^{|\kappa|} H_{\rho)}{}^{\nu)\sigma}
\nn & &
 - \frac{1}{2D(D-2)}\Big(R_{\kappa\sigma}-\frac{2}{D-2}\Lambda G_{\kappa\sigma}\Big)
  (DG_{\lambda\rho}\nabla^\kappa H^{\mu\nu\sigma} + 2G^{\mu\nu}\nabla^\kappa H_{\lambda\rho}{}^\sigma) 
\nn & &
 - \frac{1}{2D}\Big(R_{\kappa(\lambda}-\frac{2}{D-2}\Lambda G_{\kappa(\lambda}\Big)
  G^{\mu\nu}\nabla^\kappa H^\sigma{}_{|\sigma|\rho)}
\nn & &
 +\frac{1}{2}\Big(R_{\kappa(\lambda}-\frac{2}{D-2}\Lambda G_{\kappa(\lambda}\Big)
  \nabla^\kappa H^{\mu\nu}{}_{\rho)}
\nn & &
 + \frac{1}{4D}\nabla_\kappa\Big[\Big(R_{\lambda\rho}-\frac{2}{D-2}\Lambda G_{\lambda\rho}\Big)
  G^{\mu\nu}H_\sigma{}^{\sigma\kappa}\Big]
\nn & &
 - \frac{1}{4}\nabla_\kappa\Big[\Big(R_{\lambda\rho}-\frac{2}{D-2}\Lambda G_{\lambda\rho}\Big)
  H^{\mu\nu\kappa}\Big]
\nn & &
 -\frac{1}{4}\p_\kappa\Big(R-\frac{2D}{D-2}\Lambda\Big) \delta_{(\lambda}{}^{(\mu}Y^{|\kappa|\nu)}{}_{\rho)}
 +\frac{1}{4}\nabla_\kappa\Big(R_{(\lambda}{}^{(\mu}-\frac{2}{D-2}\Lambda\delta_{(\lambda}{}^{(\mu}\Big)Y^{|\kappa|\nu)}{}_{\rho)}
\nn & & 
 +\frac{1}{4}\nabla_{(\lambda}\Big(R_{|\kappa|}{}^{(\mu}-\frac{2}{D-2}\Lambda\delta_{|\kappa|}{}^{(\mu}\Big)Y^{|\kappa|\nu)}{}_{\rho)}
 -\frac{1}{4}\nabla^{(\mu}\Big(R_{\kappa(\lambda}-\frac{2}{D-2}\Lambda G_{\kappa(\lambda}\Big)Y^{|\kappa|\nu)}{}_{\rho)}
\nn & & 
 -\frac{1}{4}\nabla_\kappa\Big(R_{\sigma(\lambda}-\frac{2}{D-2}\Lambda G_{\sigma(\lambda}\Big)
  \delta_{\rho)}{}^{(\mu}Y^{|\kappa\sigma|\nu)}
 +\frac{1}{4}\nabla_\kappa\Big(R_\sigma{}^{(\mu}-\frac{2}{D-2}\Lambda\delta_\sigma{}^{(\mu}\Big)
  \delta_{(\lambda}{}^{\nu)}Y^{\kappa\sigma}{}_{\rho)}
\nn & &
 -\frac{1}{2D}\nabla_\kappa\Big(R_{\sigma(\lambda}-\frac{2}{D-2}\Lambda G_{\sigma(\lambda}\Big)
  G^{\mu\nu}Y^{\kappa\sigma}{}_{\rho)},
\label{solh0-1}
\eea
\bea
\lefteqn{
2\Big(R^{\kappa\sigma}-\frac{2}{D-2}\Lambda G^{\kappa\sigma}\Big)\nabla^{(\mu}H_{\kappa\sigma}{}^{\nu)}
} \nn & = &
 -(D-2)\nabla^{(\mu}\Big[\Big(R-\frac{2D}{D-2}\Lambda\Big)F^{\nu)\kappa}{}_\kappa\Big]
 + \frac{(D-2)^2}{2D^2}G^{\mu\nu}\nabla_\kappa\Big[\Big(R-\frac{2D}{D-2}\Lambda\Big)F^{\kappa\sigma}{}_\sigma\Big]
\nn & &
 + \frac{(D-1)(D-2)^2}{2D}\nabla_\kappa\Big(R-\frac{2D}{D-2}\Lambda\Big)H^{\mu\nu\kappa}
 + \nabla^{(\mu}\Big(R-\frac{2D}{D-2}\Lambda\Big)H_\kappa{}^{|\kappa|\nu)}
\nn & &
 - \frac{D^2-2D+2}{2D}G^{\mu\nu}\nabla_\kappa\Big(R-\frac{2D}{D-2}\Lambda\Big)H_\sigma{}^{\sigma\kappa}
\nn & &
 + \Big(R-\frac{2D}{D-2}\Lambda\Big)
\Big[
 \frac{(D-1)(D-2)(D-4)}{2D}\nabla^\kappa H^{\mu\nu}{}_\kappa
\nn & &
 +\frac{2D^2-3D+2}{D}\nabla^{(\mu}H_\kappa{}^{|\kappa|\nu)}
 -\frac{D^2+2}{2D}G^{\mu\nu}\nabla^\kappa H^\sigma{}_{\sigma\kappa}
 -\frac{D-2}{D^2}G^{\mu\nu}\nabla_\kappa H^{\kappa\sigma}{}_\sigma
\Big]
\nn & &
 +\frac{2(D-1)^2}{D}\nabla_\kappa\Big[
 \Big(R^{\mu\nu}-\frac{2}{D-2}\Lambda G^{\mu\nu}\Big)H_\sigma{}^{\sigma\kappa}\Big]
\nn & &
 -2\nabla^{(\mu}\Big(R_{\kappa\sigma}-\frac{2}{D-2}\Lambda G_{\kappa\sigma}\Big)H^{|\kappa\sigma|\nu)}
\nn & &
 -\frac{4(D-1)^2}{D}\Big(R_\kappa{}^{(\mu}-\frac{2}{D-2}\Lambda\delta_\kappa{}^{(\mu}\Big)
  \nabla^{|\kappa} H_\sigma{}^{\sigma|\nu)}
\nn & &
+2\Big(R_{\kappa\sigma}-\frac{2}{D-2}\Lambda G_{\kappa\sigma}\Big)
 \Big[ G^{\mu\nu}\nabla^\kappa H_\tau{}^{\tau\sigma} 
 +\frac{D-2}{D^2}G^{\mu\nu}\nabla^\kappa H^{\sigma\tau}{}_\tau
\nn & &
 + \frac{(D-1)(D-2)}{D}\nabla^\kappa H^{\mu\nu\sigma}
\Big]
\nn & &
 +\frac{(D-1)(D^2-6D+4)}{D}\nabla_\kappa\Big(R_\sigma{}^{(\mu}-\frac{2}{D-2}\Lambda\delta_\sigma{}^{(\mu}\Big)
  Y^{|\kappa\sigma|\nu)},
\label{solh0-2}
\eea
\bea
\lefteqn{
2\Big(R^{\kappa\sigma}-\frac{2}{D-2}\Lambda G^{\kappa\sigma}\Big)\nabla^{(\mu}H^{\nu)}{}_{\kappa\sigma}
} \nn & = & 
 \frac{2(D-2)}{D^2}G^{\mu\nu}\nabla_\kappa\Big[\Big(R-\frac{2D}{D-2}\Lambda\Big)F^{\kappa\sigma}{}_\sigma\Big]
 -\frac{D-2}{2}\nabla^{(\mu}\Big[\Big(R-\frac{2D}{D-2}\Lambda\Big)F_\kappa{}^{|\kappa|\nu)}\Big]
\nn & &
 -\frac{D-4}{2}\nabla^{(\mu}\Big[\Big(R-\frac{2D}{D-2}\Lambda\Big)H^{\nu)\kappa}{}_\kappa\Big]
\nn & &
 +\Big(R-\frac{2D}{D-2}\Lambda\Big)\Big[
 -\frac{(D-1)(D-4)}{D}\nabla^{(\mu} H_\kappa{}^{|\kappa|\nu)}
 +\frac{2(D-1)(D-4)}{D}\nabla_\kappa H^{\mu\nu\kappa}
\nn & &
 -\frac{D-4}{2D}G^{\mu\nu}\nabla_\kappa H_\sigma{}^{\sigma\kappa}
 -\frac{4}{D^2}G^{\mu\nu}\nabla_\kappa H^{\kappa\sigma}{}_\sigma
\Big]
\nn & &
 +\frac{2(D-1)(D-2)}{D}\nabla_\kappa\Big(R-\frac{2D}{D-2}\Lambda\Big)H^{\mu\nu\kappa}
\nn & &
 -\frac{3D-4}{2D}G^{\mu\nu}\nabla_\kappa\Big(R-\frac{2D}{D-2}\Lambda\Big)H_\sigma{}^{\sigma\kappa}
\nn & &
 -\frac{(D-1)(D-4)}{D}\nabla_\kappa\Big[\Big(R^{\mu\nu}-\frac{2}{D-2}\Lambda G^{\mu\nu}\Big)
 H_\sigma{}^{\sigma\kappa}\Big]
\nn & &
 +\frac{2(D-1)(D-4)}{D}\Big(R_\kappa{}^{(\mu}-\frac{2}{D-2}\Lambda\delta_\kappa{}^{(\mu}\Big)
 \nabla^{|\kappa} H_\sigma{}^{\sigma|\nu)}
\nn & &
 +\Big(R_{\kappa\sigma}-\frac{2}{D-2}\Lambda G_{\kappa\sigma}\Big)
 \Big[\frac{8}{D^2}G^{\mu\nu}\nabla^\kappa H^{\sigma\tau}{}_\tau
  -2G^{\mu\nu}\nabla^\kappa H_\tau{}^{\tau\sigma}
  +\frac{8(D-1)}{D}\nabla^\kappa H^{\mu\nu\sigma}
\Big]
\nn & &
 -2\nabla^{(\mu}\Big(R_{\kappa\sigma}-\frac{2}{D-2}\Lambda G_{\kappa\sigma}\Big)
 H^{\nu)\kappa\sigma}
\nn & &
 -\frac{(D-1)(D^2-4D+8)}{D}\nabla_\kappa\Big(R_\sigma{}^{(\mu}-\frac{2}{D-2}\Lambda\delta_\sigma{}^{(\mu}\Big)
  Y^{|\kappa\sigma|\nu)}
\nn & &
 -\frac{2(D+2)}{D}\Lambda G^{\mu\nu}\nabla_\kappa K^\kappa.
\label{solh0-3}
\eea
These also look complicated, but again Ricci tensor and $\Lambda$ appear only in the combination
of the background equations of motion, except in the last term of the last equation \eqref{solh0-3}.
So if the background equation of motion is satisfied, they are drastically simplified: 
\eqref{solh0-1} and \eqref{solh0-2} are trivially satisfied, and \eqref{solh0-3} reduces to 
\beq
\Lambda\nabla_\kappa K^\kappa=0.
\label{traceK}
\eeq
In fact this gives no further restriction on $K^\mu$ because this is satisfied by any conformal Killing vector 
with the condition $\nabla_\kappa K^\kappa=$ const. in the backgrounds satisfying the vacuum equation of motion.
(See \eqref{LieCKYRicci}.)
With \eqref{traceK} the final condition for $K^\mu$ slightly varies with the value of $\Lambda$,
and if the background equation of motion is satisfied,
the final form of the symmetry operators are given as follows:
\bea
Q_{\lambda\rho}{}^{\mu\nu}h_{\mu\nu} & = &
 \mathcal{L}_K h_{\lambda\rho}
 + Y_{(\lambda}{}^{\kappa\phi}\nabla_{|\phi|}h_{\rho)\kappa}
 + \Big(c+\frac{2}{D}\nabla_\kappa K^\kappa\Big)h_{\lambda\rho}
 + F_{\lambda\rho}{}^\mu\nabla^\nu h_{\mu\nu},
\label{solQ}
\\
S_{\lambda\rho}{}^{\mu\nu}h_{\mu\nu} & = &
 \mathcal{L}_K h_{\lambda\rho}
 + Y_{(\lambda}{}^{\kappa\phi}\nabla_{|\phi|}h_{\rho)\kappa}
 + ch_{\lambda\rho} + \nabla_{(\lambda}(H^{\mu\nu}{}_{\rho)}h_{\mu\nu}),
\label{solS}
\eea
where $c=\wt{c}-\frac{4}{D}\nabla_\kappa K^\kappa$ is a constant,
$Y_{\mu\nu\lambda}$ is a Killing-Yano 3-form, $H^{\mu\nu}{}_\lambda=H^{\nu\mu}{}_\lambda$ is an arbitrary tensor,
$F_{\lambda\rho}{}^\mu=F_{\rho\lambda}{}^\mu$ is an arbitrary tensor, and
$K^\mu$ is a conformal Killing vector satisfying the condition $\nabla_\kappa K^\kappa=$ const. if $\Lambda=0$,
or $K^\mu$ is a Killing vector if $\Lambda\neq 0$.

The above result is obtained for $D\ge 4$, but it can be confirmed that it gives a solution even for $D=3$,
although it may not be general solution.

The terms proportional to $c$ in \eqref{solQ} and \eqref{solS} correspond to \eqref{trivial0}.
The term consisting of $H^{\mu\nu}{}_\lambda$ in \eqref{solS} corresponds to \eqref{trivial1}, and
the term consisting of $F_{\lambda\rho}{}^\mu$ in \eqref{solQ} corresponds to \eqref{trivial2}.
The first terms in the right hand side of \eqref{solQ} and \eqref{solS} correspond to \eqref{symKV}. 
These are anticipated from the invariance under general coordinate transformations, but
the terms proportional to $Y_{\mu\nu\lambda}$ are somewhat unexpected, and give a commuting operator.
Some of eigenfunctions of this term may give nontrivial deformations of the background geometry, and 
correspond to the moduli of Einstein manifolds. 
Odd dimensional Kerr-NUT-(A)dS spacetime (See e.g. \cite{fkk17} and references therein.)
is an example of background manifolds admitting Killing-Yano 3-forms.
Sasakian manifolds, 6D nearly K\"ahler manifolds, 7D weak $G_2$ manifolds, and 
the sphere $S^D$ of scalar curvature $D(D-1)$ are Euclidean examples \cite{s02}.
In the case of $\Lambda=0$, nonzero constant $\nabla_\kappa K^\kappa$ gives
a noncommuting symmetry operator, and it is also interpreted as a result of general coordinate transformation, 
because if $h_{\mu\nu}$ is a solution,
$\nabla_\mu K_\nu + \nabla_\nu K_\mu + \mathcal{L}_K h_{\mu\nu}$
is also a solution, which is in the form of \eqref{gct}.
In fact, the conformal transformation with $\nabla_\kappa K^\kappa=$ const. is a constant rescaling, and 
the difference between $S_{\lambda\rho}{}^{\mu\nu}$ and $Q_{\lambda\rho}{}^{\mu\nu}$ comes from
the rescaling of the background metric in $M_{\lambda\rho}{}^{\mu\nu}$.

By analyses similar to ours given in the next section we can construct general form of higher order symmetry
operators in principle.
However such calculations become increasingly difficult as the order of the operators becomes higher and higher.
One of immediate methods to give higher order operators is to take products of first order operators. 
Although it gives no more information on the solutions than the first order operators,
it may give a hint about the general forms of higher operators.

%%%%%%%%%%%%%%%%%%%%%%%%%%%%%%%%%%%%%%%%%%%%%%%%%%%%%%%%%%%%%%%%%%%
\section{A procedure for solving the conditions for the symmetry operators}

In this section we show the details of the procedure for solving the conditions 
\eqref{Ho3}, \eqref{ho2}, \eqref{Ho1}, and \eqref{ho0}. 
Since full expressions of equations appearing at intermediate steps are often lengthy, 
we only show the algorithm for solving them and do not show unnecessary details. 
Our basic strategy is to express tensors in terms of tensors with fewer free indices.
\paragraph{Analysis of \eqref{Ho3} for $S^{\lambda\rho}{}_{\mu\nu}{}^\phi$}
First let us solve \eqref{Ho3}, which is purely algebraic and does not contain derivatives.
By contracting $\sigma$ and $\tau$ in \eqref{Ho3}, we obtain
\bea
& & D(S^{\lambda\rho}{}_{\mu\nu}{}^\phi - S_\kappa{}^\kappa{}_{\mu\nu}{}^\phi G^{\lambda\rho})
-2S^{(\lambda|\phi|}{}_{\mu\nu}{}^{\rho)}
+2S_\kappa{}^\kappa{}_{\mu\nu}{}^{(\lambda}G^{\rho)\phi}
+2S^\phi{}_{\kappa\mu\nu}{}^\kappa G^{\lambda\rho}
-2S^{(\lambda}{}_{\kappa\mu\nu}{}^{|\kappa|}G^{\rho)\phi}
\nn & & =\text{(terms proportional to $Q_{\gamma\delta}{}^{\alpha\beta\epsilon}$)}.
\label{Ho3st}
\eea
By symmetrizing the indices $\lambda$, $\rho$ and $\phi$ in this equation, we obtain
\beq
(D-2)(S^{(\lambda\rho}{}_{\mu\nu}{}^{\phi)} - S_\kappa{}^\kappa{}_{\mu\nu}{}^{(\phi} G^{\lambda\rho)})
 = \text{(terms proportional to $Q_{\gamma\delta}{}^{\alpha\beta\epsilon}$)}.
\eeq
With this equation $S^{(\lambda\rho}{}_{\mu\nu}{}^{\phi)}$ can be expressed by 
$S_\kappa{}^\kappa{}_{\mu\nu}{}^{(\phi} G^{\lambda\rho)}$ and terms proportional to $Q_{\gamma\delta}{}^{\alpha\beta\epsilon}$.
Then using it and 
\beq
S^{(\lambda|\phi|}{}_{\mu\nu}{}^{\rho)}
 = -\frac{1}{2}S^{\lambda\rho}{}_{\mu\nu}{}^\phi
 + \frac{3}{2}S^{(\lambda\rho}{}_{\mu\nu}{}^{\phi)},
\eeq
$S^{(\lambda|\phi|}{}_{\mu\nu}{}^{\rho)}$ can be expressed by $S^{\lambda\rho}{}_{\mu\nu}{}^\phi$, 
$S_\kappa{}^\kappa{}_{\mu\nu}{}^{(\phi} G^{\lambda\rho)}$ and terms proportional to $Q_{\gamma\delta}{}^{\alpha\beta\epsilon}$.
By using it we can eliminate $S^{(\lambda|\phi|}{}_{\mu\nu}{}^{\rho)}$ in \eqref{Ho3st}, and we obtain
\bea
S^{\lambda\rho}{}_{\mu\nu}{}^\phi & = & 
 S_\kappa{}^\kappa{}_{\mu\nu}{}^\phi G^{\lambda\rho}
 +\frac{1}{(D+1)}\Big[
 -2S^\phi{}_{\kappa\mu\nu}{}^\kappa G^{\lambda\rho}
 +2S^{(\lambda}{}_{\kappa\mu\nu}{}^{|\kappa|} G^{\rho)\phi}
\Big]
\nn & & 
+\text{(terms proportional to $Q_{\gamma\delta}{}^{\alpha\beta\epsilon}$)}.
\label{Ho3-0}
\eea
By contracting $\lambda$ and $\rho$ in \eqref{Ho3-0},
\bea
S_\kappa{}^\kappa{}_{\mu\nu}{}^\phi & = & \frac{2}{D+1}S^\phi{}_{\kappa\mu\nu}{}^\kappa
+\text{(terms proportional to $Q_{\gamma\delta}{}^{\alpha\beta\epsilon}$)},
\label{Ho3lr}
\eea
and the contraction $\rho=\phi$ in \eqref{Ho3-0} leads to an equation of similar (but different) form:
\bea
S_\kappa{}^\kappa{}_{\mu\nu}{}^\phi & = & \frac{2}{D+1}S^\phi{}_{\kappa\mu\nu}{}^\kappa
+\text{(terms proportional to $Q_{\gamma\delta}{}^{\alpha\beta\epsilon}$)}.
\label{Ho3rp}
\eea
The difference of \eqref{Ho3lr} and \eqref{Ho3rp} gives a relation containing only $Q_{\gamma\delta}{}^{\alpha\beta\epsilon}$.
Using it we can simplify \eqref{Ho3lr}:
\bea
S_\kappa{}^\kappa{}_{\mu\nu}{}^\phi & = & \frac{2}{D+1}S^\phi{}_{\kappa\mu\nu}{}^\kappa
\nn & &
-\frac{1}{(D-2)(D+1)}\Big[
 DQ_\kappa{}^\kappa{}_{\mu\nu}{}^\phi - 2Q_\kappa{}^{\kappa\phi}{}_{(\mu\nu)}
 -DQ_\kappa{}^\kappa{}_\sigma{}^{\sigma\phi}G_{\mu\nu}
\nn & &
 +2Q_\kappa{}^{\kappa\phi}{}_\sigma{}^\sigma G_{\mu\nu}
 +2Q_\kappa{}^\kappa{}_\sigma{}^\sigma{}_{(\mu}\delta_{\nu)}{}^\phi
 -2Q_\kappa{}^\kappa{}_{\sigma(\mu}{}^\sigma\delta_{\nu)}{}^\phi
\Big].
\eea
By using this \eqref{Ho3-0} can be simplified further:
\beq
S^{\lambda\rho}{}_{\mu\nu}{}^\phi = \frac{2}{D+1}S^{(\lambda}{}_{\kappa\mu\nu}{}^{|\kappa|}G^{\rho)\phi}
 +\text{(terms proportional to $Q_{\gamma\delta}{}^{\alpha\beta\epsilon}$)}.
\label{Ho3-1}
\eeq
Each term in the right hand side of \eqref{Ho3} can be rewritten by this, and we obtain
\bea
0 & = & \frac{1}{(D-2)(D+1)}\Big[
 2Q_{\lambda\rho\mu\nu}{}^{(\phi}G^{\sigma\tau)} + 2(D-2)(D+1)Q_{\lambda\rho}{}_{(\mu}{}^{(\tau\phi}\delta_{\nu)}{}^{\sigma)}
\nn & &
 -(D-2)(D+1)Q_{\lambda\rho}{}^{(\sigma\tau\phi)}G_{\mu\nu}
 -2(D-1)Q_{\lambda\rho}{}^{(\phi}{}_{(\mu\nu)}G^{\sigma\tau)}
 +2DQ^{(\phi}{}_{(\lambda|\mu\nu|\rho)}G^{\sigma\tau)}
\nn & &
 -4Q^{(\phi}{}_{(\lambda\rho)(\mu\nu)}G^{\sigma\tau)}
 +D(D+1)Q^{(\sigma\tau}{}_{\mu\nu}{}^{\phi)}G_{\lambda\rho}
 -2(D+1)Q^{(\sigma\tau\phi)}{}_{(\mu\nu)}G_{\lambda\rho}
\nn & &
 -2D^2Q_{(\lambda}{}^{(\sigma}{}_{|\mu\nu|}{}^\phi\delta_{\rho)}{}^{\tau)}
 +4DQ_{(\lambda}{}^{(\sigma\phi}{}_{|(\mu\nu)|}\delta_{\rho)}{}^{\tau)}
 -2DQ^{(\sigma\tau}{}_{\mu\nu(\lambda}\delta_{\rho)}{}^{\phi)}
\nn & &
 +4Q^{(\sigma\tau}{}_{(\lambda|(\mu\nu)|}\delta_{\rho)}{}^{\phi)}
\Big]
\nn & & 
 +\text{(terms proportional to $Q_{\gamma\delta}{}^{\alpha\beta\epsilon}$ with some pairs of indices contracted)}.
\label{Fo3-0}
\eea
Note that this no longer contains $S^{\gamma\delta}{}_{\alpha\beta}{}^\epsilon$.
Therefore \eqref{Ho3-1} exhaustively contains information on $S^{\gamma\delta}{}_{\alpha\beta}{}^\epsilon$.
\paragraph{Analysis of \eqref{Ho3} for $Q^{\lambda\rho}{}_{\mu\nu}{}^\phi$}
By contracting $\mu$ and $\nu$ in \eqref{Fo3-0}, we obtain
\beq
Q_{\lambda\rho}{}^{(\phi\sigma\tau)} =
 \text{(terms proportional to $Q_{\gamma\delta}{}^{\alpha\beta\epsilon}$ with some pairs of indices contracted)},
\label{Fo3-iisss}
\eeq
and contracting $\lambda$ and $\rho$ in \eqref{Fo3-0} we obtain
\bea
0 & = & DQ^{(\sigma\tau}{}_{\mu\nu}{}^{\phi)} - 2Q^{(\sigma\tau\phi)}{}_{(\mu\nu)}
\nn & & +\text{(terms proportional to $Q_{\gamma\delta}{}^{\alpha\beta\epsilon}$ with some pairs of indices contracted)}
\nn & = &
 (D+1)Q^{(\sigma\tau}{}_{\mu\nu}{}^{\phi)}
\nn & &
 - ( Q^{(\sigma\tau\phi)}{}_{(\mu\nu)} + Q^{(\sigma\tau}{}_{(\mu}{}^{\phi)}{}_{\nu)} + Q^{(\sigma\tau}{}_{\mu\nu}{}^{\phi)})
\nn & &
 +\text{(terms proportional to $Q_{\gamma\delta}{}^{\alpha\beta\epsilon}$ with some pairs of indices contracted)}.
\eea
The second line of the last expression of the above equation is symmetric under the interchange of 
the last three indices, and therefore it can be eliminated by \eqref{Fo3-iisss}. Then
\beq
Q^{(\sigma\tau}{}_{\mu\nu}{}^{\phi)} = 
 \text{(terms proportional to $Q_{\gamma\delta}{}^{\alpha\beta\epsilon}$ with some pairs of indices contracted)}.
\label{Fo3-ssiis}
\eeq
From \eqref{Fo3-iisss} and \eqref{Fo3-ssiis},
\bea
Q^{\sigma\tau\phi(\mu\nu)} & = & Q^{\phi(\sigma|\mu\nu|\tau)}
 +\frac{3}{2}(Q^{\sigma\tau(\mu\nu\phi)} - Q^{(\sigma\tau|\mu\nu|\phi)})
\nn & = & 
 Q^{\phi(\sigma|\mu\nu|\tau)}
 + \text{(terms proportional to $Q_{\gamma\delta}{}^{\alpha\beta\epsilon}$ }
\nn & &
 \text{ with some pairs of indices contracted)}.
\label{Fo3-iiiss}
\eea
Rewriting the third term of \eqref{Fo3-0} by \eqref{Fo3-iisss},
the seventh term by \eqref{Fo3-ssiis},
the fourth, sixth, eighth, tenth, and twelfth terms by \eqref{Fo3-iiiss},
we obtain
\bea
0 & = &
 Q_{\lambda\rho(\mu}{}^{(\tau}{}^\phi\delta_{\nu)}{}^{\sigma)}
 - Q_{(\lambda}{}^{(\sigma}{}_{|\mu\nu|}{}^\tau\delta_{\rho)}{}^{\phi)}
 - \frac{1}{(D-2)}Q^{(\sigma\tau}{}_{\mu\nu}{}^{\phi)} G_{\lambda\rho}
\nn & & 
 + \text{(terms proportional to $Q_{\gamma\delta}{}^{\alpha\beta\epsilon}$ with some pairs of indices contracted)}.
\eea
The third term of the above can be rewritten by \eqref{Fo3-ssiis}:
\bea
0 & = &
 Q_{\lambda\rho(\mu}{}^{(\tau}{}^\phi\delta_{\nu)}{}^{\sigma)}
 - Q_{(\lambda}{}^{(\sigma}{}_{|\mu\nu|}{}^\tau\delta_{\rho)}{}^{\phi)}
\nn & & 
 + \text{(terms proportional to $Q_{\gamma\delta}{}^{\alpha\beta\epsilon}$ with some pairs of indices contracted)}.
\eea
Contracting $\nu$ and $\phi$ in the above,
\bea
0 & = & (D+3)Q_{\lambda\rho\mu}{}^{(\sigma\tau)}
 -2Q_{(\lambda}{}^{(\sigma}{}_{|\mu|\rho)}{}^{\tau)}
\nn & &
 + \text{(terms proportional to $Q_{\gamma\delta}{}^{\alpha\beta\epsilon}$ with some pairs of indices contracted)}.
\label{Fo3-a1}
\eea
From \eqref{Fo3-iisss}
\bea
Q_{\lambda\rho}{}^{\mu(\sigma\tau)} & = &
 -\frac{1}{2}Q_{\lambda\rho}{}^{\sigma\tau\mu}
 +\frac{3}{2}Q_{\lambda\rho}{}^{(\mu\sigma\tau)}
\nn & = & 
 -\frac{1}{2}Q_{\lambda\rho}{}^{\sigma\tau\mu}
 + \text{(terms proportional to $Q_{\gamma\delta}{}^{\alpha\beta\epsilon}$ }
\nn & &
 \text{ with some pairs of indices contracted)},
\label{Fo3-a2}
\eea
and furthermore
\bea
Q_{(\lambda}{}^{(\sigma}{}_{|\mu|\rho)}{}^{\tau)} & = &
 \frac{1}{2}(Q_{(\lambda}{}^{(\sigma}{}_{|\mu|\rho)}{}^{\tau)}
 +Q^{(\sigma}{}_{(\lambda}{}_{|\mu|\rho)}{}^{\tau)}
 +Q^{\sigma\tau}{}_{\mu(\rho\lambda)}
)
\nn & &
 - \frac{1}{4}(Q^{\sigma\tau}{}_{\mu(\rho\lambda)}
 + Q^{\sigma\tau}{}_{(\rho|\mu|\lambda)}
 + Q^{\sigma\tau}{}_{\rho\lambda\mu}
)
\nn & &
 + \frac{1}{4}Q^{\sigma\tau}{}_{\rho\lambda\mu}.
\eea
Since the first line of the right hand side of the above is symmetric
under the interchange of the first, second, and fifth indices,
and the second line is symmetric under the interchange of the last three indices,
they can be eliminated by \eqref{Fo3-ssiis} and \eqref{Fo3-iisss}. Then
\bea
Q_{(\lambda}{}^{(\sigma}{}_{|\mu|\rho)}{}^{\tau)} & = & \frac{1}{4}Q^{\sigma\tau}{}_{\lambda\rho\mu}
 + \text{(terms proportional to $Q_{\gamma\delta}{}^{\alpha\beta\epsilon}$ }
\nn & &
 \text{ with some pairs of indices contracted)}.
\label{Fo3-a3}
\eea
Using \eqref{Fo3-a2} and \eqref{Fo3-a3} for \eqref{Fo3-a1}, we obtain
\bea
0 & = & (D+3)Q_{\lambda\rho}{}^{\sigma\tau}{}_\mu + Q^{\sigma\tau}{}_{\lambda\rho\mu}
\nn & &
 + \text{(terms proportional to $Q_{\gamma\delta}{}^{\alpha\beta\epsilon}$ with some pairs of indices contracted)}.
\label{Fo3-a4}
\eea
Subtracting \eqref{Fo3-a4} with $(\lambda,\rho)$ and $(\sigma,\tau)$ exchanged from \eqref{Fo3-a4} times $(D+3)$,
we obtain
\beq
Q_{\lambda\rho}{}^{\sigma\tau}{}_\mu =
 \text{(terms proportional to $Q_{\gamma\delta}{}^{\alpha\beta\epsilon}$ with some pairs of indices contracted)}.
\label{Fo3-f0}
\eeq
Thus we have succeeded in expressing $Q_{\lambda\rho}{}^{\sigma\tau}{}_\mu$ in terms of 
tensors with fewer free indices. Contractions of indices of this and other equations
lead relations between tensors with some indices contracted.
By using them we can simplify \eqref{Fo3-f0} more.

From the contractions $(\lambda,\tau)=(\rho,\phi)$ and $(\lambda,\rho)=(\tau,\phi)$
in \eqref{Fo3-iisss}, and $(\sigma,\nu)=(\tau,\phi)$ in \eqref{Fo3-ssiis},
\bea
Q_{\kappa\mu}{}^\kappa{}_\sigma{}^\sigma & = & \frac{1}{2}(D+1)Q_{\kappa\mu\sigma}{}^{\sigma\kappa},
\nn
Q_{\kappa\sigma}{}^\sigma{}_\mu{}^\kappa & = & \frac{1}{2}Q_{\kappa\mu\sigma}{}^{\sigma\kappa}
 +\frac{1}{2}Q_\kappa{}^\kappa{}_\sigma{}^\sigma{}_\mu,
\nn
Q_{\kappa\sigma}{}^{\kappa\sigma}{}_\mu & = & Q_{\kappa\mu\sigma}{}^{\sigma\kappa}
 -Q_\kappa{}^\kappa{}_{\sigma\mu}{}^\sigma
 +\frac{1}{2}(D+1)Q_\kappa{}^\kappa{}_\sigma{}^\sigma{}_\mu.
\label{Fo3-c0}
\eea
By using these, $Q_{\kappa\mu}{}^\kappa{}_\sigma{}^\sigma$, $Q_{\kappa\sigma}{}^\sigma{}_\mu{}^\kappa$, and
$Q_{\kappa\sigma}{}^{\kappa\sigma}{}_\mu$ are expressed in terms of $Q_{\kappa\mu\sigma}{}^{\sigma\kappa}$, 
$Q_\kappa{}^\kappa{}_\sigma{}^\sigma{}_\mu$, and $Q_\kappa{}^\kappa{}_{\sigma\mu}{}^\sigma$.
Contracting $\lambda$ and $\rho$ in \eqref{Ho3}, renaming $\sigma$ and $\tau$ to $\lambda$ and $\rho$,
and adding \eqref{Ho3-0} times $D-2$ with $\lambda$, $\rho$ and $\phi$ symmetrized, we obtain
\bea
0 & = & DQ_{(\lambda\rho}{}^{\mu\nu}{}_{\phi)}-2Q_{(\lambda\rho\phi)}{}^{(\mu\nu)}
\nn & & + \text{(terms proportional to $Q_{\gamma\delta}{}^{\alpha\beta\epsilon}$ with some pairs of indices contracted)}.
\label{Fo3-c1}
\eea
Contracting $\rho$ and $\phi$ in \eqref{Ho3} and rewriting 
$S^{\gamma\delta}{}_{\alpha\beta}{}^\epsilon$ with no contraction of indices in it by \eqref{Ho3-0}, we obtain
\bea
0 & = &
 (D+2)Q_{\lambda(\sigma}{}^{\mu\nu}{}_{\tau)}
 -2Q_{\lambda(\sigma}{}^{(\mu}{}_{\tau)}{}^{\nu)}
 -2Q_{\lambda}{}^{(\mu\nu)}{}_{(\sigma\tau)}
\nn & & + \text{(terms proportional to $Q_{\gamma\delta}{}^{\alpha\beta\epsilon}$ with some pairs of indices contracted)}.
\label{Fo3-c2}
\eea
From the difference of \eqref{Ho3lr} and \eqref{Ho3rp}, and the
contraction $\nu=\phi$ in \eqref{Fo3-c1} minus $\frac{3(2D-1)}{2(D-2)}$ times the same equation with the indices 
$\lambda$, $\rho$, $\mu$ symmetrized, 
\bea
Q_{\lambda\kappa}{}^{\mu\nu\kappa} & = & 
 \text{(terms proportional to 
 $Q_\kappa{}^{\kappa\alpha\beta\epsilon}$,
 $Q_{\gamma\delta\kappa}{}^{\kappa\epsilon}$,
 $Q_{\gamma\kappa}{}^{\alpha\kappa\epsilon}$,
}
\nn & &
 \text{
 $Q_{\kappa\alpha\sigma}{}^{\sigma\kappa}$, 
 $Q_\kappa{}^\kappa{}_\sigma{}^\sigma{}_\alpha$, or
 $Q_\kappa{}^\kappa{}_{\sigma\alpha}{}^\sigma$
 )},
\label{Fo3-c3}
\\
Q_{\lambda\rho}{}^{\mu\kappa}{}_\kappa & = & 
 \text{(terms proportional to 
 $Q_\kappa{}^{\kappa\alpha\beta\epsilon}$,
 $Q_{\gamma\delta\kappa}{}^{\kappa\epsilon}$,
 $Q_{\gamma\kappa}{}^{\alpha\kappa\epsilon}$,
}
\nn & &
 \text{
$Q_{\kappa\alpha\sigma}{}^{\sigma\kappa}$,
$Q_\kappa{}^\kappa{}_\sigma{}^\sigma{}_\alpha$, or
$Q_\kappa{}^\kappa{}_{\sigma\alpha}{}^\sigma$
 )}.
\label{Fo3-c4}
\eea
From the contraction $\mu=\nu$ in \eqref{Fo3-c2} with $\lambda$ renamed $\mu$, plus
the contraction $\lambda=\nu$ in \eqref{Fo3-c2} times $2(D+1)^2$, 
\beq
Q_{\mu\kappa}{}^{\kappa(\sigma\tau)} = 
  \text{(terms proportional to 
 $Q_\kappa{}^{\kappa\alpha\beta\epsilon}$,
 $Q_{\gamma\delta\kappa}{}^{\kappa\epsilon}$,
 $Q_{\kappa\alpha\sigma}{}^{\sigma\kappa}$,
 $Q_\kappa{}^\kappa{}_\sigma{}^\sigma{}_\alpha$, or
 $Q_\kappa{}^\kappa{}_{\sigma\alpha}{}^\sigma$
 )}
\label{Fo3-c5}.
\eeq
Therefore, with the definition $\phi_\mu{}^{\sigma\tau}\equiv Q_{\mu\kappa}{}^{\kappa[\sigma\tau]}$,
\bea
Q_{\mu\kappa}{}^{\kappa\sigma\tau} & = &
 Q_{\mu\kappa}{}^{\kappa[\sigma\tau]} + Q_{\mu\kappa}{}^{\kappa(\sigma\tau)}
\nn & = &
 \phi_\mu{}^{\sigma\tau}
 + \text{(terms proportional to 
 $Q_\kappa{}^{\kappa\alpha\beta\epsilon}$,
 $Q_{\gamma\delta\kappa}{}^{\kappa\epsilon}$, }
\nn & & 
 \text{$Q_{\kappa\alpha\sigma}{}^{\sigma\kappa}$,
 $Q_\kappa{}^\kappa{}_\sigma{}^\sigma{}_\alpha$, or
 $Q_\kappa{}^\kappa{}_{\sigma\alpha}{}^\sigma$
 )}.
\label{Fo3-c6}
\eea
Contracting $\mu$ and $\sigma$ in \eqref{Fo3-c6},
\beq
Q_\kappa{}^\kappa{}^{\mu\sigma}{}_\sigma =  
 \frac{1}{2}Q^{\kappa\mu\sigma}{}_{\sigma\kappa} - 2\phi_\kappa{}^{\kappa\mu}
 + \frac{D}{2}Q_\kappa{}^\kappa{}_\sigma{}^{\sigma\mu}.
\label{Fo3-c7}
\eeq
From the contraction $\lambda=\tau$ in \eqref{Fo3-c6}, minus $\frac{3}{2}\frac{3D^2+ 9D+8}{(D+2)(4D+1)}$ times
the same equation with the indices $\mu$, $\nu$ and $\sigma$ symmetrized,
\bea
Q_\kappa{}^{\kappa\mu\nu\sigma} & = &
 \text{(terms proportional to $\phi_\alpha{}^{\gamma\delta}$, $Q_{\gamma\delta\kappa}{}^{\kappa\epsilon}$,
 $\phi_\kappa{}^{\kappa\alpha}$, 
 $Q_{\kappa\alpha\sigma}{}^{\sigma\kappa}$, or
 $Q_\kappa{}^\kappa{}_\sigma{}^\sigma{}_\alpha$)}.
\label{Fo3-c8}
\eea
Contracting $\mu$ and $\rho$ in \eqref{Fo3-f0}, we obtain an equation which is proportional to $D-3$.
Since we consider $D\ge 4$, we can drop this factor (This is the only step at which we use the condition $D\ge 4$.),
and 
\bea
 Q_{\lambda\tau\kappa}{}^\kappa{}_\sigma
 + Q_{\lambda\sigma\kappa}{}^\kappa{}_\tau
 -2Q_{\sigma\tau\kappa}{}^\kappa{}_\lambda
 & = & 
 \text{(terms proportional to $\phi_\alpha{}^{\gamma\delta}$, }
\nn & & 
 \text{ $\phi_\kappa{}^{\kappa\alpha}$, $Q_{\kappa\alpha\sigma}{}^{\sigma\kappa}$, or
 $Q_\kappa{}^\kappa{}_\sigma{}^\sigma{}_\alpha$
 )}.
\label{Fo3-c9}
\eea
Then, with the definition $\Phi_{\lambda\sigma\tau}\equiv Q_{(\sigma\tau|\kappa|}{}^\kappa{}_{\lambda)}$,
\bea
Q_{\sigma\tau\kappa}{}^\kappa{}_\lambda & = & 
 Q_{(\sigma\tau|\kappa|}{}^\kappa{}_{\lambda)}
 - \frac{1}{3}(
 Q_{\lambda\tau\kappa}{}^\kappa{}_\sigma
 + Q_{\lambda\sigma\kappa}{}^\kappa{}_\tau
 -2Q_{\sigma\tau\kappa}{}^\kappa{}_\lambda
)
\nn & = & 
 \Phi_{\lambda\sigma\tau}
 + \text{(terms proportional to $\phi_\alpha{}^{\gamma\delta}$, 
 $\phi_\kappa{}^{\kappa\alpha}$, $Q_{\kappa\alpha\sigma}{}^{\sigma\kappa}$, or
 $Q_\kappa{}^\kappa{}_\sigma{}^\sigma{}_\alpha$
 )}.
\label{Fo3-c10}
\eea
Contracting $\sigma=\tau$ in \eqref{Fo3-c10}
\beq
Q_{\kappa\mu\sigma}{}^{\sigma\kappa} = 
 -\frac{1}{2}Q_\kappa{}^\kappa{}_\sigma{}^\sigma{}_\mu
 +\frac{3}{2}\Phi^\kappa{}_{\kappa\mu}.
\label{Fo3-c11}
\eeq
Using \eqref{Fo3-c0}, \eqref{Fo3-c3}, \eqref{Fo3-c4}, \eqref{Fo3-c6}-\eqref{Fo3-c8}, 
\eqref{Fo3-c10}, and \eqref{Fo3-c11},
$Q_{\lambda\kappa}{}^{\mu\nu\kappa}$,
$Q_{\lambda\rho}{}^{\mu\kappa}{}_\kappa$,
$Q_{\lambda\kappa}{}^{\kappa\mu\phi}$,
$Q_\kappa{}^{\kappa\mu\nu\sigma}$, and
$Q_{\lambda\rho\kappa}{}^\kappa{}_\phi$
are expressed in terms of 
$Q_\kappa{}^\kappa{}_\sigma{}^\sigma{}_\alpha$, 
$\Phi_{\gamma\delta\epsilon}$,  
$\phi_\alpha{}^{\gamma\delta}$,
$\Phi^\kappa{}_{\kappa\alpha}$, and 
$\phi_\kappa{}^{\kappa\alpha}$.
Then from \eqref{Fo3-f0},
\beq
S_{\lambda\rho}{}^{\mu\nu\phi} =
 \text{(terms proportional to $Q_\kappa{}^\kappa{}_\sigma{}^\sigma{}_\alpha$, 
 $\Phi_{\gamma\delta\epsilon}$,  
 $\phi_\alpha{}^{\gamma\delta}$,
 $\Phi^\kappa{}_{\kappa\alpha}$, or 
 $\phi_\kappa{}^{\kappa\alpha}$)},
\label{Fo3-f1}
\eeq
and it can be confirmed that contractions of this equation reproduce 
\eqref{Fo3-c0}-\eqref{Fo3-c11}.
In this equation $Q_\kappa{}^\kappa{}_\sigma{}^{\sigma\phi}$ appears only in the combination
$Q_\kappa{}^\kappa{}_\sigma{}^{\sigma\phi} - \Phi_\kappa{}^{\kappa\phi} + \frac{8}{3}\phi_\kappa{}^{\kappa\phi}$.
To obtain a concise expression of this equation, we define $K^\phi$, $Y_{\lambda\mu\nu}$,
and $F_{\lambda\rho}{}^\mu$ as follows:
\bea
K^\phi & = & \frac{3}{2}\frac{1}{(D-1)(D+2)}\Big(
 Q_\kappa{}^\kappa{}_\sigma{}^{\sigma\phi}
 - \Phi_\kappa{}^{\kappa\phi}
 + \frac{8}{3}\phi_\kappa{}^{\kappa\phi} \Big),
\\
Y_{\lambda\mu\nu} & = & \frac{4}{D+2}\phi_{[\lambda\mu\nu]},
\\
F_{\lambda\rho}{}^\mu & = & \Phi_{\lambda\rho}{}^\mu
 - \frac{8}{3}\phi_{(\lambda\rho)}{}^\mu
 - \frac{2}{3}(D+2)\delta_{(\lambda}{}^\mu K_{\rho)}
 + \frac{1}{3}(2D+1)G_{\lambda\rho}K^\mu.
\eea
Conversely $\phi_{\lambda\mu\nu}$ and $\Phi_{\lambda\rho\mu}$ are given by
$K^\phi$, $Y_{\lambda\mu\nu}$, and $F_{\lambda\rho}{}^\mu$ as follows:
\bea
\phi_{\lambda\mu\nu} & = & \frac{1}{4}(D+2)Y_{\lambda\mu\nu}
 - \frac{1}{2}F_{\lambda[\mu\nu]} + \frac{1}{2}(D+1)G_{\lambda[\mu}K_{\nu]},
\\
\Phi_{\lambda\rho\mu} & = & F_{(\lambda\rho\mu)} + G_{(\lambda\rho}K_{\mu)}.
\eea
Then \eqref{Fo3-f1} is rewritten as follows:
\beq
Q_{\lambda\rho}{}^{\mu\nu\phi} = 
 \delta_{(\lambda}{}^\mu\delta_{\rho)}{}^\nu K^\phi
 + \delta_{(\lambda}{}^{(\mu}Y_{\rho)}{}^{\nu)}{}^\phi
 + G^{\phi(\mu}F_{\lambda\rho}{}^{\nu)}.
\label{Fo3-f2}
\eeq
Using this, \eqref{Ho3-1} is simplified further.
Let us define $H^{\mu\nu}{}_\lambda$ as follows:
\beq
H^{\mu\nu}{}_\lambda = \frac{2}{D+1}(
 S_{\kappa\lambda}{}^{\mu\nu\kappa} - \delta_\lambda{}^{(\mu}K^{\nu)} ),
\eeq
then
\beq
S_{\lambda\rho}{}^{\mu\nu\phi} = 
 \delta_{(\lambda}{}^\mu\delta_{\rho)}{}^\nu K^\phi
 + \delta_{(\lambda}{}^{(\mu}Y_{\rho)}{}^{\nu)}{}^\phi
 + \delta^\phi{}_{(\lambda}H^{\mu\nu}{}_{\rho)}.
\label{Ho3-f0}
\eeq
It is not difficult to confirm that \eqref{Fo3-f2} and \eqref{Ho3-f0} solve \eqref{Ho3}.
Therefore \eqref{Fo3-f2} and \eqref{Ho3-f0} give the general solution to \eqref{Ho3}.
\paragraph{Analysis of \eqref{ho2}}
Next let us solve \eqref{ho2}.
Contracting $\tau$ and $\phi$,
\bea
s_{\lambda\rho}{}^{\mu\nu} & = & s_\kappa{}^{\kappa\mu\nu}G_{\lambda\rho} +
 \text{(terms proportional to $q_{\gamma\delta}{}^{\alpha\beta}$ or $\nabla_\zeta S_{\gamma\delta}{}^{\alpha\beta\epsilon}$)}.
\label{ho2-1}
\eea
Contracting $\lambda$ and $\rho$ in this equation, 
\beq
s_\kappa{}^{\kappa\mu\nu} =
 \text{(terms proportional to $q_{\gamma\delta}{}^{\alpha\beta}$ or $\nabla_\zeta S_{\gamma\delta}{}^{\alpha\beta\epsilon}$)}.
\eeq
Using this and \eqref{ho2-1}, we can express $s_{\lambda\rho}{}^{\mu\nu}$ in terms of other tensors:
\beq
s_{\lambda\rho}{}^{\mu\nu} = 
 \text{(terms proportional to $q_{\gamma\delta}{}^{\alpha\beta}$ or $\nabla_\zeta S_{\gamma\delta}{}^{\alpha\beta\epsilon}$)}.
\label{ho2-2}
\eeq
Contracting $\lambda$ and $\rho$ in \eqref{ho2}, we obtain the following equation similar to (but different from) 
\eqref{ho2-1}.
\bea
s_{\lambda\rho}{}^{\mu\nu} & = & s_\kappa{}^{\kappa\mu\nu}G_{\lambda\rho} +
 \text{(terms proportional to $q_{\gamma\delta}{}^{\alpha\beta}$ or $\nabla_\zeta S_{\gamma\delta}{}^{\alpha\beta\epsilon}$)}.
\label{ho2-3}
\eea
Taking the difference of \eqref{ho2-1} and \eqref{ho2-3}, we obtain the following equation which does not contain 
$s_{\gamma\delta}{}^{\alpha\beta}$:
\bea
q^{\tau\phi\mu\nu} & = & q^{\tau\phi\kappa}{}_\kappa G^{\mu\nu} 
 +\frac{1}{D-2}\Big[
 q_\kappa{}^{\kappa\mu\nu}G^{\tau\phi} + q_\kappa{}^{\kappa\tau\phi}G^{\mu\nu}
\nn & &
 - q_\kappa{}^{\kappa\tau(\mu}G^{\nu)\phi} - q_\kappa{}^{\kappa\phi(\mu}G^{\nu)\tau}
 + q_\kappa{}^\kappa{}_\sigma{}^\sigma (G^{\tau(\mu}G^{\nu)\phi} - G^{\mu\nu}G^{\tau\phi})
\Big]
\nn & &
 +\text{(terms proportional to $\nabla_\zeta S_{\gamma\delta}{}^{\alpha\beta\epsilon}$)}.
\label{fo2-0}
\eea
Contracting $\mu$ and $\nu$ in this equation, we obtain $q^{\tau\phi\kappa}{}_\kappa$
expressed by $q_\kappa{}^{\kappa\alpha\beta}$ and terms proportional to 
$\nabla_\zeta S_{\gamma\delta}{}^{\alpha\beta\epsilon}$. Using it \eqref{fo2-0} can be simplified further:
\beq
q^{\tau\phi\mu\nu} =
 \text{(terms proportional to $q_\kappa{}^{\kappa\alpha\beta}$ or $\nabla_\zeta S_{\gamma\delta}{}^{\alpha\beta\epsilon}$)}.
\label{fo2-1}
\eeq
From \eqref{ho2-2}, \eqref{fo2-1} and \eqref{Ho3-f0},
\bea
q_{\lambda\rho}{}^{\mu\nu} & = &
 \text{(terms proportional to $q_\kappa{}^{\kappa\alpha\beta}$, $\nabla_\epsilon K^\alpha$, or
 $\nabla_\epsilon Y_{\alpha\beta\gamma}$)},
\label{fo2-f0}
\\
s_{\lambda\rho}{}^{\mu\nu} & = &
 \text{(terms proportional to $q_\kappa{}^{\kappa\alpha\beta}$, $\nabla_\epsilon K^\alpha$, 
 $\nabla_\epsilon Y_{\alpha\beta\gamma}$, or $\nabla_\epsilon H^{\alpha\beta}{}_\gamma$)}.
\label{ho2-f0}
\eea
Then \eqref{ho2} is rewritten as follows by \eqref{fo2-f0}, \eqref{ho2-f0}, and \eqref{Ho3-f0}: 
\beq
0 = \text{(terms proportional to $q_\kappa{}^{\kappa\alpha\beta}$, $\nabla_\epsilon K^\alpha$, or
 $\nabla_\epsilon Y_{\alpha\beta\gamma}$)}.
\label{ho2-4}
\eeq
From the contraction $(\mu,\rho)=(\nu,\tau)$ in \eqref{ho2-4},
\beq
\nabla^\kappa Y_{\kappa\mu\nu}=0,
\label{ho2-c1}
\eeq
and from the contraction $(\nu,\lambda)=(\rho,\tau)$ in \eqref{ho2-4},
\beq
q_\kappa{}^{\kappa\mu\nu} = \frac{1}{D}G^{\mu\nu}q_\kappa{}^\kappa{}_\sigma{}^\sigma
 - \nabla^\mu K^\nu - \nabla^\nu K^\mu +\frac{2}{D}G^{\mu\nu}\nabla_\kappa K^\kappa.
\label{ho2-c2}
\eeq
Then from the contraction $(\nu,\mu)=(\rho,\tau)$ in \eqref{ho2-4},
\beq
\nabla^\mu K^\nu + \nabla^\nu K^\mu = \frac{2}{D}G^{\mu\nu}\nabla_\kappa K^\kappa,
\label{ho2-c3}
\eeq
which means that $K^\mu$ is a conformal Killing vector.
Contracting $\nu$ and $\rho$ in \eqref{ho2-4},
\beq
0 = \nabla_\tau Y_{\lambda\mu\phi} + \nabla_\phi Y_{\lambda\mu\tau},
\label{ho2-c4}
\eeq
which means that $\nabla_\tau Y_{\lambda\mu\phi}$ is antisymmetric in the interchange of
all of four indices i.e. $Y_{\lambda\mu\nu}$ is a Killing-Yano 3-form.
This implies \eqref{ho2-c1}, and from \eqref{ho2-c2},
\beq
q_\kappa{}^{\kappa\mu\nu} = \wt{c}G^{\mu\nu},
\eeq
where $\wt{c}=\frac{1}{D}q_\kappa{}^\kappa{}_\sigma{}^\sigma$.
Then \eqref{fo2-f0} and \eqref{ho2-f0} are rewritten as follows:
\bea
q_{\lambda\rho}{}^{\mu\nu} & = & 
 \delta_{(\lambda}{}^{(\mu}(\nabla_{\rho)}K^{\nu)} - \nabla^{\nu)}K_{\rho)})
 + \delta_{(\lambda}{}^{\mu}\delta_{\rho)}{}^{\nu}\wt{c},
\label{fo2-f1}
\\
s_{\lambda\rho}{}^{\mu\nu} & = & 
 \delta_{(\lambda}{}^{(\mu}(\nabla_{\rho)}K^{\nu)} - \nabla^{\nu)}K_{\rho)})
 + \delta_{(\lambda}{}^{\mu}\delta_{\rho)}{}^{\nu}\wt{c}
 -\frac{2}{D}\delta_{(\lambda}{}^{\mu}\delta_{\rho)}{}^{\nu}\nabla_\kappa K^\kappa
 + \nabla_{(\lambda}H^{\mu\nu}{}_{\rho)}.
\label{ho2-f1}
\eea
It can be confirmed that these solve \eqref{ho2}, and therefore these give the general solution to \eqref{ho2}.
\paragraph{Analysis of \eqref{Ho1}}
Next let us analyze \eqref{Ho1}.
Rewriting \eqref{Ho1} by \eqref{Fo3-f2}, \eqref{Ho3-f0}, \eqref{fo2-f1}, and \eqref{ho2-f1},
and using \eqref{ddCKV}, \eqref{ddKY}, and Bianchi identity $R^\lambda{}_{[\rho\mu\nu]}=0$, 
we obtain
\bea
0 & = & 
 \frac{1}{4}(\nabla^\phi\nabla_{(\lambda} - \nabla_{(\lambda} \nabla^\phi)H^{\mu\nu}{}_{\rho)}
\nn & & 
 +\frac{3}{8}(\nabla^\kappa\nabla^\phi - \nabla^\phi \nabla^\kappa)H^{\mu\nu}{}_\kappa
 +\frac{1}{2}\delta_{(\lambda}{}^\phi(\nabla_{\rho)}\nabla^\kappa - \nabla^\kappa \nabla_{\rho)})H^{\mu\nu}{}_\kappa
\nn & & 
 +\frac{1}{24}\delta_{(\lambda}{}^{(\mu} R^{|\kappa\sigma\phi|\nu)}Y_{\rho)\kappa\sigma}
 +\frac{1}{24}\delta_{(\lambda}{}^{(\mu} R^{|\kappa\sigma\phi|}{}_{\rho)}Y^{\nu)}{}_{\kappa\sigma}
\nn & & 
 -\frac{1}{24}\delta_{(\lambda}{}^{(\mu} R^{|\kappa\sigma|\nu)\phi}Y_{\rho)\kappa\sigma}
 -\frac{1}{24}\delta_{(\lambda}{}^{(\mu} R^{|\kappa\sigma|\nu)}{}_{\rho)}Y^\phi{}_{\kappa\sigma}
\nn & & 
 +\frac{1}{4}G^{\phi(\mu} R^{|\kappa\sigma|\nu)}{}_{(\lambda}Y_{\rho)\kappa\sigma}
 +\frac{1}{8}\delta_{(\lambda}{}^\phi R^{\kappa\sigma}{}_{\rho)}{}^{(\mu}Y^{\nu)}{}_{\kappa\sigma}
\nn & & 
 -\frac{1}{8}G^{\mu\nu} R^{\kappa\sigma\phi}{}_{(\lambda}Y_{\rho)\kappa\sigma}
 -\frac{1}{16}G_{\lambda\rho} R^{\kappa\sigma\phi(\mu}Y^{\nu)}{}_{\kappa\sigma}
\nn & & 
 +\frac{1}{4}R_{(\lambda}{}^{(\mu|\phi\kappa|}Y_{\rho)}{}^{\nu)}{}_\kappa
 +\frac{1}{4}R_{(\lambda}{}^{(\mu\nu)\kappa}Y^\phi{}_{\rho)\kappa}
 +\frac{1}{4}R_{(\lambda}{}^{(\mu}{}_{\rho)}{}^{|\kappa|}Y^{\nu)\phi}{}_\kappa
\nn & & +
 \text{(terms proportional to $\p_\alpha\wt{c}$, $\p_\alpha(\nabla_\kappa K^\kappa)$, 
 $R^\kappa{}_\gamma Y_{\kappa\alpha\beta}$,
 $\Lambda F_{\gamma\delta}{}^\alpha$, or $\Lambda H^{\alpha\beta}{}_\gamma$)}
\nn & &
 + \text{(terms proportional to products of $F_{\gamma\delta}{}^\alpha$}
\nn & &
 \text{and Ricci scalar, Ricci tensor, or Riemann tensor)}
\nn & &
 + \text{(terms proportional to products of $H^{\alpha\beta}{}_\gamma$}
\nn & &
 \text{and Ricci scalar, Ricci tensor, or Riemann tensor)}.
\label{Ho1-f0}
\eea
The first two lines of the above can be rewritten by using
\beq
(\nabla_\lambda\nabla_\rho - \nabla_\rho \nabla_\lambda)H^{\mu\nu}{}_\phi
= R_{\lambda\rho}{}^\mu{}_\kappa H^{\kappa\nu}{}_\phi
 + R_{\lambda\rho}{}^\nu{}_\kappa H^{\mu\kappa}{}_\phi
 + R_{\lambda\rho\phi\kappa}H^{\mu\nu\kappa},
\eeq
and by applying \eqref{RKY2} to the third line, the fourth line, and each term in the fifth and sixth lines of the above,
the Riemann tensors in those terms can be rewritten in terms of Ricci tensor.
Furthermore the seventh line vanishes by \eqref{RKY}.

After these simplifications, we obtain the followings by contracting two pairs of indices in \eqref{Ho1-f0} in various ways:
\bea
\p_\mu\wt{c} & = & \text{(
 terms proportional to $F_{\mu\kappa}{}^\kappa$, $H^\kappa{}_{\kappa\mu}$,
 or $R_{\kappa\sigma}H^{\kappa\sigma}{}_\mu$
)},
\label{Ho1-f2} \\
\p_\mu(\nabla_\kappa K^\kappa) & = & \text{(terms proportional to 
 $F_\kappa{}^{\kappa\mu}$, $F_{\mu\kappa}{}^\kappa$, $H^\kappa{}_{\kappa\mu}$, $H^{\mu\kappa}{}_\kappa$,
} \nn & &
 \text{
 $R_{\kappa\sigma}H^{\kappa\sigma}{}_\mu$, or $R_{\kappa\sigma}H^{\mu\kappa\sigma}$
)},
\label{Ho1-f3} \\
R_{\kappa\sigma}F^{\kappa\sigma\mu} & = & \text{(terms proportional to 
 $F_\kappa{}^{\kappa\mu}$, $H^{\mu\kappa}{}_\kappa$,
 or $R_{\kappa\sigma}H^{\mu\kappa\sigma}$
)},
\label{Ho1-c1} \\
R_{\kappa\sigma}F_\mu{}^{\kappa\sigma} & = & \text{(terms proportional to 
 $F_{\mu\kappa}{}^\kappa$, $H^\kappa{}_{\kappa\mu}$,
 or $R_{\kappa\sigma}H^{\kappa\sigma}{}_\mu$
)},
\label{Ho1-c2} \\
R_{\mu\kappa}F_{\kappa\sigma}{}^\sigma & = & \text{(terms proportional to 
 $F_{\mu\kappa}{}^\kappa$, $H^{\mu\kappa}{}_\kappa$,
 or $R_{\mu\kappa}H^{\kappa\sigma}{}_\sigma$
)},
\label{Ho1-c3} \\
R_{\mu\kappa}F_\sigma{}^{\sigma\kappa} & = & \text{(terms proportional to 
 $F_\kappa{}^{\kappa\mu}$, $F_{\mu\kappa}{}^\kappa$, $H^\kappa{}_{\kappa\mu}$, $H^{\mu\kappa}{}_\kappa$,
} \nn & &
 \text{
 $R_{\kappa\sigma}H^{\kappa\sigma}{}_\mu$, $R_{\kappa\sigma}H^{\mu\kappa\sigma}$,
 or $R_{\mu\kappa}H_\sigma{}^{\sigma\kappa}$
)}.
\label{Ho1-c4} 
\eea
Then from
\eqref{Ho1-f0} minus 3 times \eqref{Ho1-f0} with the indices $\mu$, $\nu$, $\phi$ symmetrized,
\bea
\lefteqn{ \Big(R^{\mu\nu} - \frac{1}{2}G^{\mu\nu}R + \Lambda G^{\mu\nu}\Big)F_{\lambda\rho}{}^\phi
} \nn & = & \text{(terms proportional to
 $\Lambda F_\kappa{}^{\kappa\alpha}$, $\Lambda F_{\alpha\kappa}{}^\kappa$,
 $RF_\kappa{}^{\kappa\alpha}$, $RF_{\alpha\kappa}{}^\kappa$, }
\nn & &
 \text{$\Lambda H^\kappa{}_{\kappa\alpha}$, $\Lambda H^{\alpha\kappa}{}_\kappa$, $\Lambda H^{\alpha\beta\gamma}$, 
 $RH^\kappa{}_{\kappa\alpha}$, $RH^{\alpha\kappa}{}_\kappa$, $RH^{\alpha\beta\gamma}$, }
\nn & &
 \text{$R_{\kappa\sigma}H^{\kappa\sigma\alpha}$, $R_{\kappa\sigma}H^{\alpha\kappa\sigma}$,
 $R^{\alpha\beta}H_{\gamma\delta}{}^\epsilon$, or 
 $R^\kappa{}_\gamma Y_{\kappa\alpha\beta}$
)}.
\label{Ho1-f1}
\eea
It can be confirmed that this, \eqref{Ho1-f2}, and \eqref{Ho1-f3} solve \eqref{Ho1-f0}, and
reproduce \eqref{Ho1-c1}-\eqref{Ho1-c4}. 
Therefore \eqref{Ho1-f1}, \eqref{Ho1-f2}, and \eqref{Ho1-f3} are equivalent to \eqref{Ho1},
and no simpler equation is derived from them.

Note that Riemann tensor (with no indices contracted) does not appear in \eqref{Ho1-f2}, \eqref{Ho1-f3},
and \eqref{Ho1-f1}, and in fact they are in the forms that Ricci tensor and $\Lambda$
appear only in the form of the background equation of motion. The results are summarized in
\eqref{solH1-1}, \eqref{solH1-2}, and \eqref{solH1-3}.
\paragraph{Analysis of \eqref{ho0}}
Next let us analyze \eqref{ho0}.
Rewriting \eqref{ho0} by \eqref{Fo3-f2}, \eqref{Ho3-f0}, \eqref{fo2-f1}, and \eqref{ho2-f1}, and using
\eqref{def:CKV}, \eqref{ddCKV}, \eqref{def:KY}, and Bianchi identities
\beq
R^\lambda{}_{[\rho\mu\nu]}=0, \quad
\nabla_{[\lambda}R_{\mu\nu]\sigma\tau}=0,
\quad \nabla^\kappa R_{\kappa\lambda\mu\nu}=2\nabla_{[\mu}R_{\nu]\lambda},
\quad \nabla^\kappa R_{\kappa\mu}=\frac{1}{2}\p_\mu R,
\eeq
we obtain
\bea
0 & = & 
 \frac{1}{4}\Big[
 \nabla_{(\lambda}(\nabla_{\rho)}\nabla_\kappa-\nabla_{|\kappa|}\nabla_{\rho)})
 +(\nabla_{(\lambda}\nabla_{|\kappa|}-\nabla_\kappa\nabla_{(\lambda})\nabla_{\rho)}
\Big] H^{\mu\nu\kappa}
\nn & &
 +\frac{1}{4}\nabla^\kappa(\nabla_\kappa\nabla_{(\lambda}-\nabla_{(\lambda}\nabla_{|\kappa|})
 H^{\mu\nu}{}_{\rho)}
\nn & & 
 +\frac{1}{8}G_{\lambda\rho}\Big[
 (\nabla_\kappa\nabla_\sigma-\nabla_\sigma\nabla_\kappa)\nabla^\sigma
 +2\nabla^\sigma(\nabla_\kappa\nabla_\sigma-\nabla_\sigma\nabla_\kappa)
\Big] H^{\mu\nu\sigma}
\nn & &
 +\frac{1}{4}\nabla^{(\mu}R^{\nu)}{}_{(\lambda}{}^{\kappa\sigma}Y_{\rho)\kappa\sigma}
 +\frac{1}{8}\nabla_{(\lambda}R_{\rho)}{}^{(\mu|\kappa\sigma|}Y^{\nu)}{}_{\kappa\sigma}
\nn & &
 +\frac{3}{8}R^{\kappa\sigma}{}_{(\lambda}{}^{(\mu}\nabla_{|\kappa}Y_{\sigma|\rho)}{}^{\nu)}
\nn & &
 -\frac{1}{8}G_{\lambda\rho}G^{\mu\nu}K^\kappa\nabla_\kappa R
 +\frac{1}{4}\delta_{(\lambda}{}^\mu\delta_{\rho)}{}^\nu K^\kappa\nabla_\kappa R
\nn & &
 +\frac{1}{4}G_{\lambda\rho}K^\kappa\nabla_\kappa R^{\mu\nu}
 +\frac{1}{4}G^{\mu\nu}K^\kappa\nabla_\kappa R_{\lambda\rho}
 -\delta_{(\lambda}{}^{(\mu}K^{|\kappa|}\nabla_{|\kappa|} R_{\rho)}{}^{\nu)}
\nn & &
 -\frac{1}{4}G_{\lambda\rho}(R_\kappa{}^\mu\nabla^{[\kappa}K^{\nu]}
  +R_\kappa{}^\nu\nabla^{[\kappa}K^{\mu]})
 -\frac{1}{4}G^{\mu\nu}(R^\kappa{}_\lambda\nabla_{[\kappa}K_{\rho]}
  +R^\kappa{}_\rho\nabla_{[\kappa}K_{\lambda]})
\nn & &
 +\frac{1}{2}\delta_{(\lambda}{}^\mu R_{\rho)\kappa}\nabla^{[\kappa}K^{\nu]}
 +\frac{1}{2}\delta_{(\lambda}{}^\nu R_{\rho)\kappa}\nabla^{[\kappa}K^{\mu]}
 +\frac{1}{2}\delta_\lambda{}^{(\mu} R^{\nu)\kappa}\nabla_{[\kappa}K_{\rho]}
 +\frac{1}{2}\delta_\rho{}^{(\mu} R^{\nu)\kappa}\nabla_{[\kappa}K_{\lambda]}
\nn & &
 + \text{(terms proportional to  $\nabla_\alpha\nabla_\beta\wt{c}$, 
 $\Lambda\nabla_\epsilon H^{\alpha\beta\gamma}$,
 $R_{\gamma\delta\zeta\eta}\nabla_\epsilon H^{\alpha\beta\theta}$,
 $\nabla_\epsilon R_{\gamma\delta\zeta\eta}H^{\alpha\beta\theta}$,
} \nn & & 
 \text{
 $\nabla_\epsilon RH^{\alpha\beta\gamma}$,
 $R_{\gamma\delta}\nabla_\epsilon H^{\alpha\beta\theta}$,
 $\nabla_\epsilon R_{\gamma\delta}H^{\alpha\beta\theta}$,
} \nn & & 
 \text{
 $R\nabla_\kappa K^\kappa$,  $\Lambda\nabla_\kappa K^\kappa$,
 $R_{\alpha\beta}\nabla_\kappa K^\kappa$,  $\nabla_\alpha\nabla_\beta\nabla_\kappa K^\kappa$, 
 $\nabla_\epsilon R_{\zeta\eta}Y_{\alpha\beta\gamma}$, $\nabla_\epsilon RY_{\alpha\beta\gamma}$
)}.
\label{ho0-1}
\eea
The first, second and third lines of the above can be simplified by replacing 
the commutators of covariant derivatives by Riemann tensors, and 
the fourth line is rewritten as follows:
\bea
\lefteqn{ \frac{1}{4}\nabla^{(\mu}R^{\nu)}{}_{(\lambda}{}^{\kappa\sigma}Y_{\rho)\kappa\sigma}
 +\frac{1}{8}\nabla_{(\lambda}R_{\rho)}{}^{(\mu|\kappa\sigma|}Y^{\nu)}{}_{\kappa\sigma}
} \nn & = & 
 \nabla^{(\mu}\Big[\frac{1}{4}R^{|\kappa\sigma|\nu)}{}_{(\lambda}Y_{\rho)\kappa\sigma}\Big]
 +\nabla_{(\lambda}\Big[\frac{1}{8}R^{\kappa\sigma}{}_{\rho)}{}^{(\mu}Y^{\nu)}{}_{\kappa\sigma}\Big]
\nn & &
 -\frac{1}{4}R^{(\nu}{}_{(\lambda}{}^{|\kappa\sigma|}\nabla^{\mu)}Y_{\rho)\kappa\sigma}
 -\frac{1}{8}R_{(\rho}{}^{(\mu|\kappa\sigma|}\nabla_{\lambda)}Y^{\nu)}{}_{\kappa\sigma}.
\eea
The first line of the above can be simplified by \eqref{RKY2}, and
the second line can be simplified by using \eqref{def:KY}:
\bea
\lefteqn{ \frac{1}{4}\nabla^{(\mu}R^{\nu)}{}_{(\lambda}{}^{\kappa\sigma}Y_{\rho)\kappa\sigma}
 +\frac{1}{8}\nabla_{(\lambda}R_{\rho)}{}^{(\mu|\kappa\sigma|}Y^{\nu)}{}_{\kappa\sigma}
} \nn & = & 
 \nabla^{(\mu}\Big[-\frac{1}{2}R^{|\kappa|}{}_{(\lambda}Y^{\nu)}{}_{\rho)\kappa}\Big]
 +\nabla_{(\lambda}\Big[-\frac{1}{4}R^{\kappa(\mu}Y_{\rho)}{}^{\nu)}{}_{\kappa}\Big]
\nn & &
 -\frac{1}{4}R^{(\nu}{}_{(\lambda}{}^{|\kappa\sigma|}\nabla_{|\kappa}Y_{\sigma|}{}^{\mu)}{}_{\rho)}
 -\frac{1}{8}R_{(\rho}{}^{(\mu|\kappa\sigma|}\nabla_{|\kappa}Y_{\sigma|\lambda)}{}^{\nu)}
\nn & = &
 -\frac{1}{2}\nabla^{(\mu}R^{|\kappa|}{}_{(\lambda}Y^{\nu)}{}_{\rho)\kappa}
 -\frac{1}{4}\nabla_{(\lambda}R^{\kappa(\mu}Y_{\rho)}{}^{\nu)}{}_{\kappa}
\nn & &
 -\frac{3}{8}R_{(\lambda}{}^{(\mu|\kappa\sigma|}\nabla_{|\kappa}Y_{\sigma|\rho)}{}^{\nu)}.
\eea
The last term of the above cancels the fifth line of \eqref{ho0-1}.
The sixth and seventh line are simplified by the following relations derived from \eqref{LieCKY}:
\bea
K^\kappa \nabla_\kappa R_{\mu\nu} & = & 
 -\nabla_\mu K^\kappa R_{\kappa\nu} -\nabla_\nu K^\kappa R_{\kappa\mu}
\nn & &
 -\frac{D-2}{D}\nabla_\mu\nabla_\nu\nabla_\kappa K^\kappa
 -\frac{1}{D}G_{\mu\nu}\nabla^\sigma\nabla_\sigma\nabla_\kappa K^\kappa
\nn & = &
 -\nabla_{[\mu}K_{\kappa]}R^\kappa{}_\nu -\nabla_{[\nu}K_{\kappa]}R^\kappa{}_\mu
 - \frac{2}{D}\nabla_\kappa K^\kappa R_{\mu\nu}
\nn & &
 -\frac{D-2}{D}\nabla_\mu\nabla_\nu\nabla_\kappa K^\kappa
 -\frac{1}{D}G_{\mu\nu}\nabla^\sigma\nabla_\sigma\nabla_\kappa K^\kappa,
\\
K^\kappa \p_\kappa R & = &
 -\frac{2(D-1)}{D}\nabla^\sigma\nabla_\sigma\nabla_\kappa K^\kappa
 - \frac{2}{D}\nabla_\kappa K^\kappa R,
\label{LieCKYRicci}
\eea
where we used 
\beq
\nabla_\mu K_\nu = 
\nabla_{[\mu}K_{\nu]} + \frac{1}{D}G_{\mu\nu}\nabla_\kappa K^\kappa.
\eeq
Then terms proportional to $\nabla_{[\mu}K_{\nu]}$ from the seventh line cancel the eighth and ninth lines of \eqref{ho0-1}.
 
\eqref{ho0-1} is further simplified by \eqref{solH1-1} and \eqref{solH1-2}.
Then from the contraction $(\mu,\lambda)=(\nu,\rho)$ and $(\mu,\nu)=(\lambda,\rho)$ in \eqref{ho0-1},
\bea
\nabla_\kappa H^{\kappa\sigma\tau}R_{\sigma\tau} & = &
 \text{( terms proportional to 
 $\nabla_\kappa F_\sigma{}^{\sigma\kappa} \Lambda$,
 $\nabla_\kappa F^{\kappa\sigma}{}_\sigma \Lambda$,
 $\nabla_\kappa F_\sigma{}^{\sigma\kappa} R$,
 $\nabla_\kappa F^{\kappa\sigma}{}_\sigma R$,
}
\nn & &
 \text{
 $F^{\kappa\sigma}{}_\sigma \nabla_\kappa R$,
 $F_\sigma{}^{\sigma\kappa}\nabla_\kappa R$,
}
\nn & &
 \text{
 $\nabla_\kappa H_\sigma{}^{\sigma\kappa} \Lambda$, 
 $\nabla_\kappa H^{\kappa\sigma}{}_\sigma \Lambda$,
 $\nabla_\kappa H_\sigma{}^{\sigma\kappa} R$,
 $\nabla_\kappa H^{\kappa\sigma}{}_\sigma R$, 
 $H^{\kappa\sigma}{}_\sigma\nabla_\kappa R$,
 $H_\sigma{}^{\sigma\kappa}\nabla_\kappa R$,
 }
\nn & &
 \text{
 $\nabla_\kappa H_{\sigma\tau}{}^\tau R^{\kappa\sigma}$,
 $H^{\sigma\tau\kappa}\nabla_\sigma R_{\tau\kappa}$,
 $\nabla_\kappa H^\tau{}_{\tau\sigma}R^{\kappa\sigma}$,
 $\Lambda\nabla_\kappa K^\kappa$
)},
\label{ho0-c1}
\\
\nabla_\kappa H^{\sigma\tau\kappa}R_{\sigma\tau} & = &
 \text{( terms proportional to 
 $\nabla_\kappa F^{\kappa\sigma}{}_\sigma \Lambda$,
 $\nabla_\kappa F^{\kappa\sigma}{}_\sigma R$,
 $F^{\kappa\sigma}{}_\sigma \nabla_\kappa R$,
}
\nn & &
 \text{
 $\nabla_\kappa H_\sigma{}^{\sigma\kappa} \Lambda$, 
 $\nabla_\kappa H^{\kappa\sigma}{}_\sigma \Lambda$,
 $\nabla_\kappa H_\sigma{}^{\sigma\kappa} R$,
 $\nabla_\kappa H^{\kappa\sigma}{}_\sigma R$, 
 $H_\sigma{}^{\sigma\kappa}\nabla_\kappa R$,
 }
\nn & &
 \text{
 $\nabla_\kappa H_{\sigma\tau}{}^\tau R^{\kappa\sigma}$, 
 $H^{\sigma\tau\kappa}\nabla_\kappa R_{\sigma\tau}$,
 $\nabla_\kappa H^\tau{}_{\tau\sigma}R^{\kappa\sigma}$
)}.
\label{ho0-c2}
\eea
Then using the above, from the contraction $\mu=\nu$ and $\lambda=\rho$ in \eqref{ho0-1},
\bea
\nabla_{(\mu}H^{\kappa\sigma}{}_{\nu)}R_{\kappa\sigma} & = & 
 \text{( terms proportional to 
 $\Lambda\nabla_\kappa F^{\kappa\sigma}{}_\sigma$,
 $\Lambda\nabla_\alpha F_{\beta\kappa}{}^\kappa$,
}
\nn & &
 \text{
 $R\nabla_\kappa F^{\kappa\sigma}{}_\sigma$,
 $R\nabla_\alpha F_{\beta\kappa}{}^\kappa$,
 $F^{\kappa\sigma}{}_\sigma\nabla_\kappa R$,
 $F_{\beta\kappa}{}^\kappa\nabla_\alpha R$,
}
\nn & &
 \text{
 $\Lambda\nabla^\alpha H_\kappa{}^{\kappa\beta}$,
 $\Lambda\nabla_\kappa H_\sigma{}^{\sigma\kappa}$,
 $\Lambda\nabla_\kappa H^{\kappa\sigma}{}_\sigma$,
 $\Lambda\nabla_\kappa H^{\alpha\beta\kappa}$,
}
\nn & &
 \text{
 $R\nabla^\alpha H_\kappa{}^{\kappa\beta}$,
 $R\nabla_\kappa H_\sigma{}^{\sigma\kappa}$,
 $R\nabla_\kappa H^{\kappa\sigma}{}_\sigma$,
 $R\nabla_\kappa H^{\alpha\beta\kappa}$,
 }
\nn & &
 \text{
 $H_\sigma{}^{\sigma\kappa}\nabla_\kappa R$,
 $H_\kappa{}^{\kappa\alpha}\nabla^\beta R$,
 $H^{\alpha\beta\kappa}\nabla_\kappa R$,
 }
\nn & &
 \text{
 $R_{\alpha\beta}\nabla_\kappa H_\sigma{}^{\sigma\kappa}$,
 $R_\kappa{}^\alpha\nabla^\kappa H_\sigma{}^{\sigma\beta}$,
 $R_{\kappa\sigma}\nabla^\kappa H_\tau{}^{\tau\sigma}$,
 $R_{\kappa\sigma}\nabla^\kappa H^{\sigma\tau}{}_\tau$,
 $R_{\kappa\sigma}\nabla^\kappa H^{\alpha\beta\sigma}$,
 }
\nn & &
 \text{
 $H_\sigma{}^{\sigma\kappa}\nabla_\kappa R_{\alpha\beta}$,
 $H^{\kappa\sigma\alpha}\nabla^\beta R_{\kappa\sigma}$,
 }
\nn & &
 \text{ $\nabla_\kappa R_{\sigma\alpha}Y_\beta{}^{\kappa\sigma}$
)},
\label{ho0-c3}
\\
\nabla_{(\mu}H_{\nu)}{}^{\kappa\sigma}R_{\kappa\sigma} & = & 
 \text{( terms proportional to 
}
\nn & &
 \text{
 $\Lambda\nabla^\alpha F_\kappa{}^{\kappa\beta}$,
 $\Lambda\nabla_\kappa F^{\kappa\sigma}{}_\sigma$,
 $R\nabla^\alpha F_\kappa{}^{\kappa\beta}$,
 $R\nabla_\kappa F^{\kappa\sigma}{}_\sigma$,
 $F_\kappa{}^{\kappa\alpha} \nabla^\beta R$,
 $F^{\kappa\sigma}{}_\sigma \nabla_\kappa R$,
} \nn & &
 \text{
 $\Lambda\nabla^\alpha H^{\beta\kappa}{}_\kappa$,
 $\Lambda\nabla^\alpha H_\kappa{}^{\kappa\beta}$,
 $\Lambda\nabla_\kappa H_\sigma{}^{\sigma\kappa}$,
 $\Lambda\nabla_\kappa H^{\kappa\sigma}{}_\sigma$,
 $\Lambda\nabla_\kappa H^{\alpha\beta\kappa}$,
}
\nn & &
 \text{
 $R\nabla^\alpha H^{\beta\kappa}{}_\kappa$,
 $R\nabla^\alpha H_\kappa{}^{\kappa\beta}$,
 $R\nabla_\kappa H_\sigma{}^{\sigma\kappa}$,
 $R\nabla_\kappa H^{\kappa\sigma}{}_\sigma$,
 $R\nabla_\kappa H^{\alpha\beta\kappa}$,
 }
\nn & &
 \text{
 $H_\sigma{}^{\sigma\kappa}\nabla_\kappa R$,
 $H^{\alpha\kappa}{}_\kappa\nabla^\beta R$,
 $H^{\alpha\beta\kappa}\nabla_\kappa R$,
 }
\nn & &
 \text{
 $R_{\alpha\beta}\nabla_\kappa H_\sigma{}^{\sigma\kappa}$,
 $R_\kappa{}^\alpha\nabla^\kappa H_\sigma{}^{\sigma\beta}$,
 $R_{\kappa\sigma}\nabla^\kappa H^{\sigma\tau}{}_\tau$,
 $R_{\kappa\sigma}\nabla^\kappa H_\tau{}^{\tau\sigma}$,
 $R_{\kappa\sigma}\nabla^\kappa H^{\alpha\beta\sigma}$,
 }
\nn & &
 \text{
 $H_\sigma{}^{\sigma\kappa}\nabla_\kappa R_{\alpha\beta}$,
 $H^{\alpha\kappa\sigma}\nabla^\beta R_{\kappa\sigma}$,
 }
\nn & &
 \text{ $\nabla_\kappa R_{\sigma\alpha}Y_\beta{}^{\kappa\sigma}$,
 $\Lambda\nabla_\kappa K^\kappa$
)}.
\label{ho0-c4}
\eea
\eqref{ho0-1} simplified by the above, \eqref{ho0-c3}, and \eqref{ho0-c4} are equivalent to \eqref{ho0-1}.
They can be rewritten in the form given in \eqref{solh0-1}, \eqref{solh0-2}, and \eqref{solh0-3}.
No simpler relation can be derived from them.

%%%%%%%%%%%%%%%%%%%%%%%%%%%%%%%%%%%%%%%%%%%%%%%%%%%%%%%%%%%%%%
\vs{.5cm}
\noindent
{\large\bf Acknowledgments}\\[.2cm]
I would like to thank T.~Houri for correspondence.
%%%%%%%%%%%%%%%%%%%%%%%%%%%%%%%%%%%%%%%%%%%%%%%%%%%%%%%%%%%%%%

\renewcommand{\theequation}{\Alph{section}.\arabic{equation}}
\appendix
\addcontentsline{toc}{section}{Appendix}
%%%%%%%%%%%%%%%%%%%%%%%%%%%%%%%%%%%%%%%%%%%%%%%%%%%%%%%%%%%%%
\vs{.5cm}
\noindent
{\Large\bf Appendix}
\section{Conformal Killing vectors and Killing-Yano tensors}
\label{appa}
\setcounter{equation}{0}

A conformal Killing vector $K^\mu$ is defined by
\beq
\nabla_\mu K_\nu + \nabla_\nu K_\mu = \frac{2}{D}g_{\mu\nu}\nabla_\lambda K^\lambda.
\label{def:CKV}
\eeq
If the right hand side vanishes, $K^\mu$ is a Killing vector.
From this equation we can show the following:
\beq
\nabla_\mu\nabla_\nu K_\lambda = 
 R_{\lambda\nu\mu\rho}K^\rho
 + \frac{1}{D}\Big[g_{\lambda\mu}\p_\nu(\nabla_\rho K^\rho)
 + g_{\lambda\nu}\p_\mu(\nabla_\rho K^\rho)
 - g_{\mu\nu}\p_\lambda(\nabla_\rho K^\rho) \Big],
\label{ddCKV}
\eeq
and the Lie derivative of Riemann tensor along $K^\mu$ is given by
\beq
\mathcal{L}_KR^\lambda{}_{\rho\mu\nu} =
 \frac{1}{D}(\delta_\nu{}^\lambda\delta_\rho{}^\sigma
 -g_{\rho\nu}g^{\lambda\sigma})\nabla_\mu\nabla_\sigma\nabla_\tau K^\tau -
 \frac{1}{D}(\delta_\mu{}^\lambda\delta_\rho{}^\sigma
 -g_{\rho\mu}g^{\lambda\sigma})\nabla_\nu\nabla_\sigma\nabla_\tau K^\tau.
\label{LieCKY}
\eeq

A Killing-Yano 3-form $Y_{\lambda_1\lambda_2\lambda_3}$ is defined as an antisymmetric tensor obeying
\beq
\nabla_\mu Y_{\lambda_1\lambda_2\lambda_3} = \nabla_{[\mu}Y_{\lambda_1\lambda_2\lambda_3]}.
\label{def:KY}
\eeq
From this equation we can show the following:
\beq
\nabla_\mu\nabla_\nu Y_{\lambda_1\lambda_2\lambda_3}
 = -2R_{\mu\rho[\nu\lambda_1}Y^\rho{}_{\lambda_2\lambda_3]},
\label{ddKY}
\eeq
\beq
0 = R_{\mu\sigma_1[\nu}{}^\rho Y_{|\rho|\sigma_2\lambda]}
 +R_{\nu\sigma_1[\mu}{}^\rho Y_{|\rho|\sigma_2\lambda]}
 +(\sigma_1\leftrightarrow\sigma_2).
\label{RKY}
\eeq
The following relation derived from the above is also useful:
\beq
R_{\rho\sigma\lambda(\mu}Y^{\rho\sigma}{}_{\nu)} = 2R_{\rho(\mu}Y^\rho{}_{\nu)\lambda}.
\label{RKY2}
\eeq

%%%%%%%%%%%% References %%%%%%%%%%%%%%%%%%%%%%%%%
\newcommand{\J}[4]{{\sl #1} {\bf #2} (#3) #4}
\newcommand{\andJ}[3]{{\bf #1} (#2) #3}
\newcommand{\AP}{Ann.\ Phys.\ (N.Y.)}
\newcommand{\MPL}{Mod.\ Phys.\ Lett.}
\newcommand{\NP}{Nucl.\ Phys.}
\newcommand{\PL}{Phys.\ Lett.}
\newcommand{\PR}{Phys.\ Rev.}
\newcommand{\PRL}{Phys.\ Rev.\ Lett.}
\newcommand{\PTP}{Prog.\ Theor.\ Phys.}
\newcommand{\hepth}[1]{{\tt hep-th/#1}}
\newcommand{\arxivhep}[1]{{\tt arXiv.org:#1 [hep-th]}}
%%%%%%%%%%%%%%%%%%%%%%%%%%%%%%%%%%%%%%%%%%%%%%%%

%%%%%%%%%%%%%%%%%%%%%%%%%%%%%%%%%%%%%%%%%%%%%%%%%%%%%%%%%%%%%%%%%%%
\end{document}